\documentclass[a4paper,fleqn]{cas-sc}

\usepackage[numbers]{natbib}
\usepackage{cleveref}
\usepackage{lipsum}
\usepackage{physics}
\usepackage{tikz}
\newcommand{\Real}[1]{\mathrm{Real}\{#1\}}
\newcommand{\Pe}{\mathrm{Pe}}
\newcommand{\Nu}{\mathrm{Nu}}
\renewcommand{\Re}{\mathrm{Re}}
\renewcommand{\Pr}{\mathrm{Pr}}

\usepackage{mathtools}
\usepackage{atveryend}
\DeclarePairedDelimiter{\avg}{\langle}{\rangle}

\let\tmp\phi
\let\phi\varphi
\let\varphi\tmp
\renewcommand{\exp}[1]{\mathrm{exp}(#1)}

\usepackage{subcaption}
\usepackage{siunitx}

\usepackage{environ}
\NewEnviron{optional}[1]{%
	\ifnum#1=1\relax
	\BODY
	\else
	\huge\texttt{<currently commented out>}%
	\fi
}
\usepackage{pgfplots}
\usepackage{tikz}
\usepackage{tikz-3dplot}
\usepgfplotslibrary{groupplots} 
\usetikzlibrary{external}
\usetikzlibrary{patterns}
\usepgflibrary{patterns.meta}
\usetikzlibrary{arrows}
\usetikzlibrary{arrows.meta}
\usetikzlibrary{calc}
\usetikzlibrary{decorations.text}
\usepgfplotslibrary{colorbrewer}
\usepgfplotslibrary{polar}
\usepgfplotslibrary{fillbetween} 
\usepackage{pgfplotstable}
\usepackage{mathtools}
\usetikzlibrary{
	decorations.markings
}
\pgfplotsset{compat = 1.18,
	invoke before crossref tikzpicture={\tikzexternaldisable},
	invoke after crossref tikzpicture={\tikzexternalenable},
}

\makeatletter
\newcommand\gpsubtitle[1]{\footnotesize (\@alph{\pgfplots@group@current@plot}) #1}
\makeatother
\tikzsetnextfilename{sketch_channel}
\tikzset{lss/.style={fill=none}, hss/.style={fill=black!30}}
\tdplotsetmaincoords{80}{-70}

\tikzset{->-/.style = {decoration={markings,
			mark=at position 0.5 with {\arrow{stealth}}},
		postaction={decorate},
		, draw=black!70
	},}

\tikzset{
	every node/.style={
		font=\footnotesize
	}
}

\pgfdeclaredecoration{example}{initial}
{
	\state{initial}[width=0.2cm]%
	{%
		\pgfpatharc{0}{180}{0.1cm}%
	}%
	\state{final}
	{
		\pgfpathmoveto{\pgfpointdecoratedpathlast}
	}
}

\makeatletter
\let\pgfplots@orig@preparenewcontourline=\pgfplotsplothandlercontour@preparenewcontourline

\pgfplotsset{
	contour line style/.initial=,
}

\def\pgfplotsplothandlercontour@preparenewcontourline@styled{%
	\pgfplots@orig@preparenewcontourline
	\pgfkeysgetvalue{/pgfplots/contour line style}\pgfplots@loc@style
	\ifx\pgfplots@loc@style\pgfutil@empty
	\else
	\tikzset{\pgfplots@loc@style}%
	\fi

	\pgfkeysgetvalue{/pgfplots/contour level threshold}\pgfplots@loc@thresh
	\ifx\pgfplots@loc@thresh\pgfutil@empty
	\def\pgfplots@loc@thresh{0.4}%
	\fi
	\pgfplotscoordmath{meta}{parsenumber}{\pgfplots@loc@thresh}%
	\let\pgfplots@contourthreshold=\pgfmathresult
	\pgfplotscoordmath{meta}{if less than}%
	{\pgfplots@current@point@meta}{\pgfplots@contourthreshold}{%
		\pgfsetlinewidth{0.2pt}%
	}{%
		\pgfsetlinewidth{0.8pt}%
	}%
}

\pgfplotsset{
	contour level style on/.code={
		\let\pgfplotsplothandlercontour@preparenewcontourline
		=\pgfplotsplothandlercontour@preparenewcontourline@styled
	},
	contour level style off/.code={
		\let\pgfplotsplothandlercontour@preparenewcontourline
		=\pgfplots@orig@preparenewcontourline
	},
	contour level threshold/.initial=0.4,
}
\pgfplotsset{contour level style off}
\makeatother

\makeatletter
\let\pgfplots@orig@preparenewcontourline=\pgfplotsplothandlercontour@preparenewcontourline

\pgfplotsset{
	contour line style/.initial=red,
}

\def\pgfplotsplothandlercontour@preparenewcontourline{%
	\pgfplots@orig@preparenewcontourline
	\pgfkeysgetvalue{/pgfplots/contour line style}\pgfplots@loc@style
	\ifx\pgfplots@loc@style\pgfutil@empty
	\else
	\tikzset{\pgfplots@loc@style}%
	\fi
}
\makeatother

\newcommand{\drawsubplotlabel}[1]{
\node[text width=1em,anchor=north west, yshift=0.15cm] at (rel axis cs:0,1)
{\subcaption{\label{#1}}};
}

\pgfplotsset{
subplotlabel/.code={%
	\edef\axissubplotlabel{#1}%
	\pgfkeysalso{
		execute at begin axis={%
			\drawsubplotlabel{#1}
		}%
	}%
},
}

\pgfplotsset{
subplotlabel at end/.code={%
	\pgfkeysalso{
		execute at end axis={%
			\drawsubplotlabel{#1}
		}%
	}%
},
}

\pgfplotsset{legend style={nodes={scale=0.625, transform shape}},}

\colorlet{MBCcolor}        {black!20!green}
\colorlet{K0.25_L1.0color} {blue!50!green}
\colorlet{K1.0_L1.0color}  {blue}
\colorlet{K4.0_L1.0color}  {red!50!blue}
\colorlet{IFcolor}         {red}
\colorlet{K1.0_L0.5color}  {orange}
\colorlet{K1.0_L2.0color}  {brown}

\tikzset{
MBC/.style = {dashed, MBCcolor}, %
K0.25_L1.0/.style = {K0.25_L1.0color, mark=none},
K1.0_L0.5/.style = {K1.0_L0.5color, mark=none},
K1.0_L1.0/.style = {K1.0_L1.0color, mark=none},
K1.0_L2.0/.style = {K1.0_L2.0color, mark=none},
K4.0_L1.0/.style = {K4.0_L1.0color, mark=none},
IF/.style = {dashed, IFcolor} %
}
\tikzset{
Straub_MBC/.style = {only marks, black, mark=o, mark size=1pt, mark repeat=15},
Straub_IF/.style = {only marks, black, mark=x, mark size=1pt, mark repeat=15},
}

\tikzset{axesfillsmall/.pic={
\draw[fill=black!50 ] (rel axis cs:1.06, 0.25) rectangle  (rel axis cs:1,0); 
\draw[] (rel axis cs:1,0.25) rectangle (rel axis cs:1.06, 0.75);
\draw[fill=black!50] (rel axis cs:1,0.75) rectangle (rel axis cs:1.06, 1);
}}

\pgfplotsset{%
my legend/.style={legend image code/.code={%
\node[##1] at (0cm,0cm){\pgfuseplotmark{triangle}};
\node[##1] at (0.15cm,0cm){\pgfuseplotmark{*}};
\node[##1] at (0.3cm,0cm){\pgfuseplotmark{diamond}};
\node[##1] at (0.45cm,0cm){\pgfuseplotmark{square*}};
\path[draw=none, use as bounding box] (0cm, -0.1cm) rectangle (0.65cm, 0.1cm);
}},%
}

\usepackage{etoolbox}
\usepackage{framed}
\usepackage[noprefix]{nomencl}
\makenomenclature

\renewcommand\nomgroup[1]{%
  \item[\bfseries
  \ifstrequal{#1}{A}{Latin symbols}{%
  \ifstrequal{#1}{B}{Greek symbols}{%
  \ifstrequal{#1}{C}{Subscripts}{%
  \ifstrequal{#1}{D}{Superscripts and modifiers}{%
  \ifstrequal{#1}{E}{Abbreviations}{%
  }}}}}
]}
\newcommand{\nomlatin}[1][]{\nomenclature[A#1]}
\newcommand{\nomgreek}[1][]{\nomenclature[B#1]}
\newcommand{\nomsubs}[1][]{\nomenclature[C#1]}
\newcommand{\nomsuper}[1][]{\nomenclature[D#1]}
\newcommand{\nomabbr}[1][]{\nomenclature[E#1]}

\begin{document}
\let\WriteBookmarks\relax
\def\floatpagepagefraction{1}
\def\textpagefraction{.001}
\shorttitle{Conjugate heat transfer in turbulent pipe flow with azimuthally non-uniform heating}
\shortauthors{J. Neuhauser et~al.}

\title [mode = title]{Conjugate heat transfer in turbulent pipe flow with azimuthally non-uniform heating}                      

\author[1]{Jonathan Neuhauser}[
                        auid=000,
                        orcid=0000-0003-3288-6056]
\cormark[1]
\fnmark[1]
\ead{jonathan.neuhauser@kit.edu}

\credit{Conceptualization, Methodology, Software, Validation, Formal analysis, Investigation, Data Curation, Visualization, Writing -- Original draft}

\affiliation[1]{organization={Institute of Fluid Mechanics, Karlsruhe Institute of Technology},
                addressline={Kaiserstr. 12}, 
                city={Karlsruhe},
                postcode={76131}, 
                country={Germany}}

\author[1]{Davide Gatti}[orcid=0000-0002-8178-9626]

\credit{Conceptualization, Methodology, Resources, Supervision,  Writing -- Review \& Editing}

\author[1]{Bettina Frohnapfel}[orcid=0000-0002-0594-7178]

\credit{Conceptualization, Supervision, Resources, Project administration, Funding acquisition,  Writing -- Review \& Editing}

\begin{abstract}
Turbulent conjugate heat transfer (CHT) in pipes with azimuthally non-uniform heating is studied theoretically and numerically with a focus on low-$\Pr$ fluids.  While the influence of the solid properties is well documented for uniform heat flux, their effect under azimuthally non-uniform heating has not been explored systematically. We address this gap by combining spectral analysis of the problem -- which reduces arbitrary boundary conditions to a discrete set of azimuthal wavenumbers -- with well-resolved numerical simulations of thermally and hydraulically fully developed turbulent pipe flow at $\Re_\tau \approx 180$ for $\Pr = \{0.025, 0.71\}$, representative of liquid metals and air. Two principal results emerge: (i) the global Nusselt number is independent of the azimuthal heat-flux distribution and exhibits only a minor dependence on the solid properties, resolving an open question for low-Prandtl-number fluids; (ii) CHT nevertheless strongly affects the azimuthally resolved mean temperature at low $\Pr$, where azimuthal conduction in the solid homogenises the interface heat flux, and governs the level of temperature fluctuations through the effusivity ratio (thermal activity ratio) $K$. Finally, an analysis of the turbulent Prandtl number $\Pr_{t}$ shows a strong anisotropy, particularly near the wall, and a dependence on the azimuthal wavenumber.
\end{abstract}

\begin{graphicalabstract}
\end{graphicalabstract}

\begin{highlights}
\item CHT has strong impact on local mean temperature for low-$\Pr$ and non-uniform heating 
\item CHT has little influence on azimuthally averaged temperature statistics and $\Nu$
\item Spectral analysis of the problem reduces parameter space to azimuthal wavenumbers
\item Turbulent Prandtl number is anisotropic and depends on shape of boundary condition
\end{highlights}


\maketitle

\begin{framed}
\nomabbr{$\Pr$}{Prandtl number, $\Pr = \nu\rho_f c_{p,f} / \lambda_f$}
\nomabbr{$\Pr_{t}$}{Turbulent Prandtl number, $\Pr_{t,i} = \nu_t / \alpha_{t,i}$ (\cref{eq:def_alphat}), $i \in \{r,\phi\}$}
\nomabbr{$\Pe$}{Péclet number, $\Pe = \Re_b \Pr$ (\cref{sec:equations})}
\nomabbr{$\Re$}{Reynolds number}
\nomabbr{$\Nu$}{Nusselt number (\cref{eq:nusselt_definition})}
\nomabbr{$\Nu_l$}{Local Nusselt number (\cref{eq:nulocal_def})}

\nomabbr{CFL}{Courant--Friedrichs--Lewy number}
\nomabbr{CHT}{Conjugate heat transfer}
\nomabbr{LES}{Large-eddy simulation}
\nomabbr{DNS}{Direct numerical simulation}
\nomabbr{IF}{Isoflux boundary condition (prescribed fluid heat flux, no solid)}
\nomabbr{MBC}{Mixed boundary condition (prescribed mean heat flux with inhibited wall temperature fluctuations)}
\nomabbr{CSP}{Concentrated solar power}
\nomabbr{$\Real{\cdot}$}{Real part of complex number}

\nomlatin{$K$}{Ratio of effusivities (thermal activity ratio), $K = \sqrt{\lambda_f \rho_f c_{p,f} / \lambda_s \rho_s c_{p,s}}$}
\nomlatin{$\vec u, u_i$}{Velocity vector / component}
\nomlatin{$u_r, u_\phi, u_z$}{Radial, azimuthal and streamwise velocity components}
\nomlatin{$u_b$}{Bulk velocity, $u_b = 1/A \int_A \avg{u_z} \dd A$}
\nomlatin[$u_\tau$]{$u_\tau$}{Friction velocity}
\nomlatin{$r$}{Radial coordinate}
\nomlatin{$z$}{Streamwise coordinate}
\nomlatin{$t$}{Time}
\nomlatin{$\Phi$}{Generic (arbitrary) field}
\nomlatin{$p$}{Pressure}
\nomlatin{$T$}{Temperature}
\nomlatin{$T_p$}{Instantaneous temperature of phase $p$ (\cref{tempeqn})}
\nomlatin{$T_w$}{Azimuthally and statistically averaged interface temperature, $T_w = \overline{\avg{T}}_{R_i}$}
\nomlatin{$T_b$}{Bulk temperature, $T_b = 1/(A u_b) \int_A \avg{u_z T_f} \dd A$}
\nomlatin{$T_{b,l}$}{Local bulk temperature of a given azimuthal slice (\cref{eq:nusselt_local_definition})}
\nomlatin{$q(\phi)$}{Heating function applied at the outer wall, $q(\phi) \in L^2$}
\nomlatin{$q_w(\phi)$}{Wall heat flux}
\nomlatin[$q_w$]{$q_{w,\mathrm{ref}}$}{Reference heat flux (\cref{eq:qref})}
\nomlatin{$f(\phi)$}{Dimensionless heat flux distribution, $f(\phi) = q_w(\phi) / q_{w,\mathrm{ref}}$ (\cref{eq:nondim})}
\nomlatin{$f_k$}{$k$-th complex Fourier coefficient of $f(\phi)$ (\cref{eq:bc-fft})}
\nomlatin{$R_i$}{Radius of the pipe's inner wall (fluid--solid interface)}
\nomlatin{$R_o$}{Radius of the pipe's outer wall}
\nomlatin{$A$}{Pipe cross-sectional area}
\nomlatin{$L_z$}{Streamwise (axial) domain length}
\nomlatin{$N$}{Polynomial order (spectral element method)}
\nomlatin{$c_p$}{Specific heat capacity at constant pressure}
\nomlatin{$j$}{Imaginary unit}
\nomlatin{$k$}{Azimuthal wavenumber}
\nomlatin{$m, n$}{Exponents of velocity / temperature in a statistical moment (\cref{stat_moment})}
\nomlatin[$L$]{$L^2$}{Space of square-integrable functions}
\nomlatin[$L$]{$\mathcal{L}$}{Linear (rotation-equivariant) response operator of first-order temperature moments (\cref{eq:linearoperator_firstordermoment})}
\nomlatin{$p(u_r, u_\phi, u_z, \vartheta, \dots)$}{Joint probability density function of velocity and temperature}
\nomlatin{$E$}{Streamwise power spectral density of temperature, $E = \hat\vartheta \hat\vartheta^*$}
\nomlatin{$k_z$}{Streamwise wavenumber}
\nomlatin[$P$]{$P_{\vartheta'\vartheta'}$}{Production of temperature variance (\cref{eq:model_production_of_tt})}
\nomlatin{$nel_{cs,f}$, $nel_{cs,s}$, $nel_z$}{Number of fluid (solid) elements per cross section / elements in streamwise direction (\cref{tab:pipe_cases_metadata})}
\nomlatin{$t_{avg}$}{Averaging time (\cref{tab:pipe_cases_metadata})}

\nomgreek{$\nu$}{Kinematic viscosity}
\nomgreek{$\nu_t$}{Turbulent eddy viscosity (Boussinesq approximation, \cref{sec:cht_fdev_nonuniform})}
\nomgreek{$\rho$}{Density}
\nomgreek[$\alpha_t$]{$\alpha_{t}$}{Turbulent thermal diffusivity; $\alpha_{t,r}$, $\alpha_{t,\phi}$: radial / azimuthal component (\cref{eq:def_alphat})}
\nomgreek{$\lambda$}{Thermal conductivity}
\nomgreek{$\lambda_z$}{Streamwise wavelength (spectra)}
\nomgreek[$\lambda_sf$]{$\lambda_{sf}$}{Solid/fluid ratio of thermal conductivities, $\lambda_{sf} = \lambda_s / \lambda_f$}
\nomgreek{$\Gamma$}{Radius ratio, $\Gamma = R_o / R_i$}
\nomgreek{$\phi$}{Azimuthal coordinate}
\nomgreek{$\vartheta$}{Non-dimensional temperature}
\nomgreek[$\delta_v$]{$\delta_v$}{Viscous length scale, $\delta_v = \nu / u_\tau$}

\nomsubs{$\dottedsquare_{p}$}{Phase ($p$ being fluid $f$ or solid $s$)}
\nomsubs{$\dottedsquare_{f}$}{Fluid}
\nomsubs{$\dottedsquare_{s}$}{Solid}
\nomsubs{$\dottedsquare_{i}$}{Component index ($i \in \{r, \phi, z\}$)}
\nomsubs{$\dottedsquare_b$}{Bulk quantity}
\nomsubs{$\dottedsquare_w$}{Wall quantity}
\nomsubs{$\dottedsquare_l$}{Local quantity (of an azimuthal slice)}
\nomsubs{$\dottedsquare_k$}{Quantity associated with azimuthal wavenumber $k$}
\nomsubs{$\dottedsquare_t$}{Turbulent quantity (e.g.\ $\nu_t$, $\alpha_t$, $\Pr_t$)}
\nomsubs{$\dottedsquare_\tau$}{Friction-velocity-based quantity ($u_\tau$, $\Re_\tau$)}

\nomsuper{$\dottedsquare^*$}{Non-dimensionalised quantity (\cref{eq:nondim})}
\nomsuper{$\dottedsquare^*$}{Complex conjugate (in the context of spectra, $E = \hat\vartheta \hat\vartheta^*$)}
\nomsuper{$\dottedsquare^{+}$}{Quantity in viscous (inner) units, e.g.\ $y^+$, $\vartheta^+$ (\cref{sec:numerical-setup})}
\nomsuper{$\dottedsquare'$}{Fluctuating part (deviation from the statistical average)}
\nomsuper{$\widehat{\dottedsquare}$}{Fourier coefficient (azimuthal: index $k$; streamwise: $k_z$)}
\nomsuper{$\overline{\dottedsquare}$}{azimuthal average, $\overline{g} = \frac{1}{2\pi}\int_0^{2\pi} g \dd\phi$}
\nomsuper[$\avg$]{$\avg{\cdot}$}{Statistical (ensemble) average; approximated by averaging over time and homogeneous directions}

\printnomenclature
\end{framed}

\section{Introduction}
\begin{figure}
	\centering 
	\tikzexternaldisable
	\begin{tikzpicture}[scale=0.8]
		
		\node[anchor=north east, inner sep=0pt] (a) at (-1.5, 1) {\includegraphics[trim={1.5cm 0cm 1.5cm 0cm},clip, height=5.76cm]{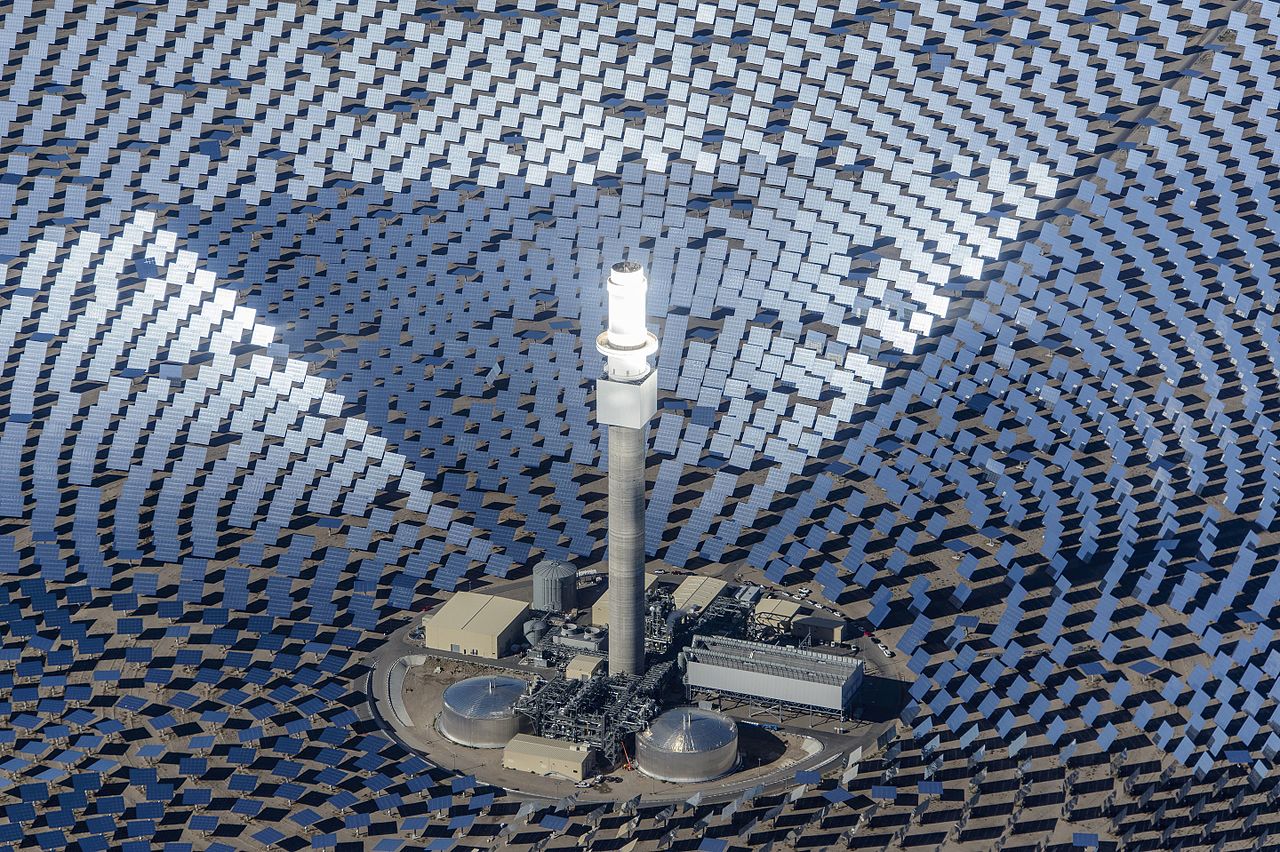}};
		\draw (-1, 1) coordinate (b) rectangle (7, -6.2);
		
		\draw[fill=orange] (0, 0) ++ (0:3.75cm and 2.8cm) arc (0:-80:3.75cm and 2.8cm) -- ++(0, -3) arc (-80:0:3.75cm and 2.8cm) -- cycle;
		\draw[fill=orange] (0, 0) ++ (0:3.75cm and 2.8cm) arc (0:-80:3.75cm and 2.8cm) -- ++(-80:-0.25cm) arc (-80:0:3.4cm and 2.55cm) -- cycle;
		\foreach[count=\i] \y in {<-, <-, ->, ->, <-, <-, ->, ->, <-, <-} {
			
			\fill[white] ({4-8*\i}:4cm and 3cm) coordinate (c\i) ++(-0.2cm, 0) -- ++(0, -3cm) -- ++(0.4cm, 0) -- ++(0, 3cm); 
			\draw[] (c\i) ++(-0.2cm, 0) -- ++(0, -3cm) ++(0.4cm, 0) -- ++(0, 3cm); 
			
			\draw[fill=white] (c\i) ellipse (0.2cm and 0.15cm);
			\draw[fill=white] (c\i) ++ (0.2cm, -3cm) arc (0:-180:0.2cm and 0.15cm);
			
			\draw[\y] (c\i) ++(0, -1.25) -- ++(0, -0.5);
			
		}
		\draw[->] (c1) -- ++(0, 0.6) node[right, align=left] {heat transfer \\ fluid};
		
		\foreach \y in {0, 0.33, ..., 1} {
			\draw[yellow!90!black, thick, <-] (4.5-\y, -4-\y) -- ++ (-30:1.5);
		}
		\path (5.5, -5.3) node[draw=yellow!90!black, thick, align=center, fill=white] {solar \\ radiation};
		
		\path [decorate,decoration={text along path,
			text={|\footnotesize|Insulation}}]
		(0, 0) ++ (-60:3.2cm and 2.4cm) arc (-60:0:3.2cm and 2.4cm) -- cycle;
		
		\draw[<-] (c10) ++ (-0.2cm, -1.5) to[out=180, in=90] ++(-0.7, -0.1) node[anchor=north east, xshift=0.5cm, align=right] {absorber \\ tubes};	
		\draw (a.north east)  ++ (-4, -2.5) coordinate (c) rectangle ++(0.2, -0.3) ;
		\draw (c) ++ (0.2, 0) -- (-1, 1);
		\draw (c) ++ (0.2, -0.3) -- (-1, -6.2);
		
		\node[text width=1em,anchor=north west, yshift=0.15cm, fill=white, fill opacity=0.5, text opacity=1] at (a.north west)
		{\subcaption{\label{fig:crescentdunes}}};
		
		\node[text width=1em,anchor=north west, yshift=0.15cm, text opacity=1] at (-1,1)
		{\subcaption{\label{fig:inside_receiver}}};
	\end{tikzpicture}
	\caption{(a) Aerial photo of the Crescent Dunes solar power plant in Nevada, United States. Image by user Marygrikas on Wikimedia (CC-BY-SA 4.0). (b) Sketch of the receiver surface. A heat transfer fluid flows through the absorber tubes, which are heated by the solar radiation from one side. The flow direction may differ between pipes.}
	\label{fig:csp}
	\tikzexternalenable
\end{figure}
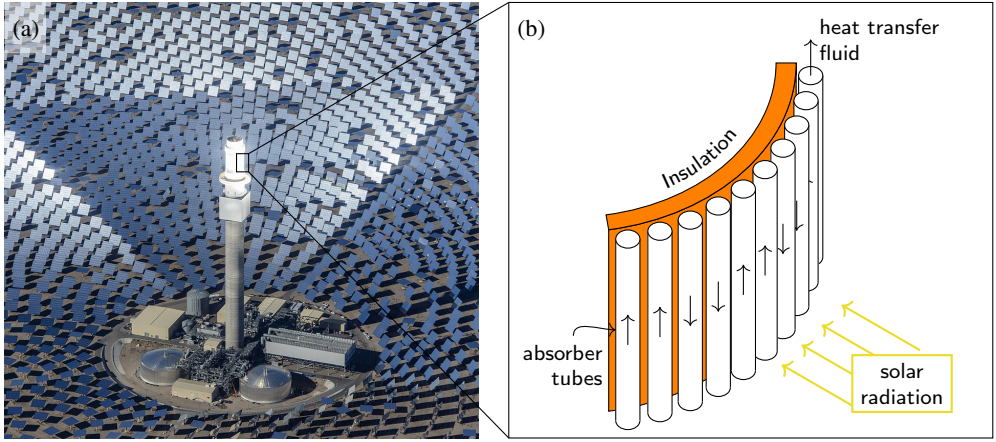

In typical heat-transfer devices, the fluid is separated from the heat source or sink by a solid wall. Therefore, the temperature fluctuations in the fluid are coupled to the unsteady heat conduction in the solid. Such a coupled situation is referred to as conjugate heat transfer (CHT). This study investigates the effects of CHT in systems with spatially non-uniform boundary conditions. In practice, it is difficult to find a heat-transfer device in which the wall temperature or heat flux is uniform: heaters and coolers often have strong spatial variations owing to non-uniform heating elements or radiation. In the present work, we consider non-uniform boundary conditions in the azimuthal direction of turbulent pipe flow in the thermally and hydraulically fully developed (and steady) state, i.e.\ far downstream of any axial changes in boundary condition. Such a setup has recently received renewed attention in the context of concentrated solar power (CSP, \citet{yangNumericalSimulationStudy2012}), where solar irradiation heats only part of the circumference of the absorber pipes inside the central receiver (see \cref{fig:csp}).

Both conjugate heat transfer analysis and spatially non-uniform boundary conditions have a significant research history. %
In the simplest case -- laminar flow, fully developed state and  uniform heat flux -- CHT does not affect the temperature profile. Fully developed \textit{turbulent} flow with uniform heat flux is a standard setup. While the presence of the solid is a necessity in experimental investigations, most numerical heat transfer studies idealise it with an effective boundary condition on the fluid. These boundary conditions are either prescribed (fluid) heat flux (Neumann boundary condition),  prescribed mean heat flux with inhibited temperature fluctuations on the wall (mixed boundary condition, MBC, see \citet{pillerDirectNumericalSimulation2005})  or Robin boundary conditions \citep{flageulDNSTurbulentChannel2015}. Solid-resolving studies (e.g. \citet{tiseljDNSTurbulentHeat2001} for channel flow and \citet{vicentecruzVersatileImmersedBoundary2023} for pipe flow) show that temperature statistics are well captured by the idealised boundary condition, provided a suitable one is chosen depending on the solid-fluid pairing and wall thickness \citep{flageulDNSTurbulentChannel2015, straubInfluenceThermalBoundary2019}. Idealised boundary conditions and solid-resolving CHT simulations show that the (fully developed) mean temperature profile in the fluid is only weakly sensitive to the solid properties under spatially uniform heat flux conditions \citep{pillerDirectNumericalSimulation2005,straubInfluenceThermalBoundary2019,vicentecruzVersatileImmersedBoundary2023}, and consequently the same conclusion can be drawn for $\Nu$.  %

All well-resolved numerical studies on conjugate heat transfer in turbulent flows to date consider uniform thermal boundary conditions on the solid. The influence of azimuthally \textit{non-uniform} thermal boundary conditions (applied to the outer solid surface) on the coupled solid-fluid system has not yet been explored systematically. While there are experimental studies -- which necessarily include the solid -- on azimuthally non-uniform heating available \citep{laubeTurbulentHeatTransfer2022}, the solid's properties are determined by the experimental setup and no systematic variation is carried out.

(Fluid) thermal boundary conditions with non-uniformity perpendicular to the mean flow direction were first considered by \citet{reynoldsTurbulentHeatTransfer1963} in fully developed pipe flow. %
\citet{reynoldsTurbulentHeatTransfer1963} derived the temperature field in laminar flow subject to azimuthally non-uniform wall heat flux, using a Fourier decomposition in the azimuthal direction. This analysis was later extended to conjugate heat transfer by \citet{reynoldsEffectWallHeat1963}. 

To arrive at a first estimate of the temperature field in turbulent flow with azimuthally non-uniform boundary conditions, \citet{reynoldsTurbulentHeatTransfer1963, sparrowTurbulentHeatTransfer1963} used a turbulent thermal diffusivity model, which was assumed to be isotropic, i.e.\ equal radial and azimuthal components. On the basis of experiments, \citet{blackExperimentsTurbulentHeat1967} argued that isotropy of the turbulent thermal diffusivity $\alpha_t$ is a reasonable approximation in the pipe core, however near the wall, its azimuthal component is much larger than the radial one, i.e.\ $\alpha_{t,\phi} \gg \alpha_{t,r}$. This anisotropy was later confirmed numerically by \citet{antoranzperalesNumericalStudyTurbulent2017} based on DNS data, %
and by \citet{straubAzimuthallyInhomogeneousThermal2019} with well-resolved simulations up to %
$\Re_\tau \approx 1000$. However, it is still unclear how the azimuthal and radial turbulent thermal diffusivities depend on the imposed azimuthal heat flux distribution, as both studies considered only one or two specific patterns, and whether conjugate heat transfer modifies these distributions for azimuthally non-uniform boundary conditions. The present work addresses this open question by systematically varying the solid properties and the spectral content of the boundary condition, which is shown to be representative of \textit{any} (sufficiently regular) boundary condition in the passive-scalar case.

Global heat transfer performance under non-uniform heat flux has been investigated both experimentally and numerically. Within their modelling assumption of an isotropic eddy diffusivity model,\citet{reynoldsTurbulentHeatTransfer1963,sparrowTurbulentHeatTransfer1963} showed that the global Nusselt number $\Nu$ is independent of the azimuthal variation of wall heat flux: `the average heat-transfer performance is equal to that of the axisymmetric heat-transfer situation.' \citep{sparrowTurbulentHeatTransfer1963}. Later studies raised doubts about the applicability of this conclusion to low-Prandtl-number fluids (e.g.\ liquid metals), arguing that these fluids obey a different heat transfer regime \citep{maroccoNumericalAnalysisSolar2016}. However, all subsequent numerical and experimental investigations \citep{maroccoNumericalAnalysisSolar2016, straubAzimuthallyInhomogeneousThermal2019, laubeTurbulentHeatTransfer2022} have consistently reported that $\Nu$ remains independent of the azimuthal heat flux distribution. Such an equivalence would simplify the thermal analysis significantly, as existing correlations could be used unchanged. In \citet{iTi2025}, we have previously provided a theoretical rationale that $\Nu$ must indeed be independent of the azimuthal heat flux distribution under passive scalar conditions based on a symmetry argument: all first-order moments in temperature must be azimuthally mean-free for a single-wavenumber heat flux distribution such as $\sin(\phi)$. In the present work we develop this argument further to greatly reduce the numerical effort -- as including conjugate heat transfer already adds three new dimensions to the parameter space -- and to simplify the data analysis. 

This paper is structured as follows. \Cref{sec:equations} introduces the governing equations and their non-dimensionalisation; and discusses how conjugate heat transfer expands the parameter space. To make this parameter space computationally tractable for the case of azimuthally non-uniform heat flux, \cref{sec:spectral_viewpoint} extends the analysis of \cite{iTi2025} towards a more general spectral structure of temperature statistics. This not only reduces the parameter space introduced by the boundary condition non-uniformity, but also greatly simplifies the numerical setup and analysis of the numerical data, and  yields $\Nu$ equivalence as a corollary. After briefly introducing the numerical setup in \cref{sec:numerical-setup},   \cref{sec:results} presents selected numerical statistics for conjugate heat transfer, particularly mean temperature, temperature variance, spectra and an analysis of the radial and azimuthal turbulent Prandtl number, and how they depend on the solid's material properties and the shape of the boundary condition.

\section{Governing equations and non-dimensionalisation}
\label{sec:equations}

Hydraulically and thermally fully developed pipe flow is considered under turbulent flow conditions. The conservation equations of mass, momentum and energy for a constant-property Newtonian fluid with temperature modelled as a passive scalar read
\begin{subequations}
\begin{align}
	\div{\vec u} = 0 \\ \pdv{\vec u}{t} + (\vec u \vdot \grad) \vec u = -\frac{1}{\rho} \grad{p} + \nu \laplacian{\vec u} \label{eq:momentum}\\
	\rho c_p \left(\pdv{T}{t} + (\vec u \vdot \grad) T\right) = \lambda \laplacian{T} 
\end{align}
\end{subequations}
with velocity $\vec u$, temperature $T$, kinematic viscosity $\nu$, density $\rho$, specific heat capacity at constant pressure $c_p$ and thermal conductivity $\lambda$. The statistically steady state is considered, e.g.\ $\avg{\pdv*{\Phi}{t}} = 0$ for any field $\Phi$. No mean in-plane velocity components ($\avg{u_r}$, $\avg{u_\phi}$) are present. Here and in \cref{sec:spectral_viewpoint}, $\avg{\Phi}$ indicates statistical averaging; in \cref{sec:results} these are approximated by averaging numerical data over time and applicable homogeneous directions. 

\newcommand{\pp}{p}

The heat transfer problem in a thick-walled pipe is set up as follows. Constant material properties for the fluid (index `f') and solid (index `s') phases are assumed, respectively. Heating through viscous dissipation and radiation are neglected. Under these conditions, the conservation equation for the (instantaneous) temperature $T_p$ ($\pp$ / phase being fluid $f$ or solid $s$) is given in cylindrical coordinates for each phase by
\begin{equation}
	\rho_\pp c_{p,\pp} \left(\pdv{T_\pp}{t} + u_r \pdv{T_\pp}{r} + \frac{u_\phi}{r} \pdv{T_\pp}{\phi} + u_z \pdv{T_\pp}{z}\right)
	= \lambda_{\pp} \left[\frac 1r \pdv{r} \left(r \pdv{T_\pp}{r} \right) + \frac{1}{r^2} \pdv[2]{T_\pp}{\phi} + \pdv[2]{T_\pp}{z} \right], \label{tempeqn}
\end{equation}
where the streamwise, radial and azimuthal coordinates are denoted with $z, r$ and $\phi$, respectively.
A boundary condition prescribing the wall-normal heat flux density is applied to the outer perimeter of the solution domain: $\left.\pdv*{T_s}{r}\right|_{R_o} = q(\phi)$, i.e. a heating function $q(\phi) \in L^2$ applied at the outer radius $R_o$. The fluid-solid interface is coupled by prescribing continuous temperature and heat flux at the interface $r=R_i$:
\begin{subequations}
	\begin{align}
		\lambda_s \left. \pdv{T_s}{r}\right|_{R_i} &= \lambda_f \left. \pdv{T_f}{r}\right|_{R_i} \quad &\text{for all $z\in\mathbb R$ and $\phi\in[0,2\pi]$}, \label{ic-dimensional-1} \\
		T_f (R_i,\phi,z) &= T_s (R_i,\phi,z) \quad &\text{for all $z\in\mathbb R $ and $\phi\in[0,2\pi]$}. \label{ic-dimensional-2}
	\end{align} \label{ic-dimensional}%
\end{subequations}
A suitable non-dimensionalisation is given in outer units using the azimuthally and statistically averaged interface temperature $T_w = \overline{\avg{T}}_{R_i}$ (the overbar $\overline{\cdot}$ denotes azimuthal averaging), the inner diameter $2R_i$, the bulk velocity $u_b = 1/A \int_A \avg{u_z} \dd A$ and fluid properties as
\newcommand{\qwref}{q_{w,\text{ref}}}
\begin{align}
	\vartheta_\pp =  \frac{\rho_f c_{p,f} u_b}{\Gamma \qwref} (T_w - T_\pp), 
	\quad u_i^* = \frac{u_i}{u_b}, 
	\quad r^* = \frac{r}{2R_i}, 
	\quad z^* = \frac{z}{2R_i}, 
	\quad t^* = \frac{u_b}{2R_i} t, \\
	f(\phi) = \frac{q_w(\phi)}{\qwref}, \quad \lambda^* = \frac{\lambda_\pp}{\lambda_f} =: \lambda_{sf}, \quad \rho^* = \frac{\rho_\pp}{\rho_f}, \quad c_{p, \pp}^* = \frac{c_{p, \pp}}{c_{p, f}}. \label{eq:nondim}
\end{align}
where $\Gamma = R_o / R_i$ is a radius ratio. The reference heat flux $q_{w,\mathrm{ref}}$ is given by \citep{iTi2025} as
\begin{equation}
	q_{w,\mathrm{ref}} = \begin{cases}
		\overline{q_w} & \overline{q_w} \neq 0, \\
		\sqrt{\frac{1}{\pi} \int_0^{2\pi} q_w(\phi)^2 \dd \phi} & \overline{q_w} = 0. 
	\end{cases}\label{eq:qref}
\end{equation}
This gives a meaningful reference heat flux for both zero-mean and non-zero mean heating functions. 

We consider the fully developed state, which is asymptotically obtained far downstream any streamwise change of the thermal boundary condition. In this state axial (turbulent and molecular) heat fluxes are constant and thus do not contribute to an axis-aligned, cylindrical control volume. The streamwise temperature gradient $\pdv*{T_\pp}{z}$ is constant in this regime, and can be computed a priori from a global energy balance. The non-dimensionalisation of $T_\pp$ is chosen to remove the linear temperature increase in $z$, and leads to an additional source term. The dimensionless temperature equation then reads, omitting star superscripts on velocity components and spatial/temporal coordinates for readability,
\begin{equation}
	\rho_\pp^* c_{p,\pp}^* \left(\pdv{\vartheta_\pp}{t} + u_r \pdv{\vartheta_\pp}{r} + \frac{u_\phi}{r} \pdv{\vartheta_\pp}{\phi} + u_z \pdv{\vartheta_\pp}{z} + 4 u_z \overline{f} \right)  = \lambda_\pp^* \frac{1}{\Pe}  \left[\frac 1{r} \pdv{r} \left(r \pdv{\vartheta_\pp}{r} \right) + \frac{1}{r^2} \pdv[2]{\vartheta_\pp}{\phi} +  \pdv[2]{\vartheta_\pp}{z}\right] .\label{eq:dimensionless_temperature}
\end{equation}
using the Péclet number $\Pe = (2 R_i u_b \rho_f c_{p, f})/(\lambda_f)$, which is the product of the bulk Reynolds number $\Re_b = 2 u_b R_i / \nu$ and the Prandtl number $\Pr = \nu\rho_f c_{p,f} / \lambda_f$, i.e.\ $\Pe = \Re_b \Pr$. The boundary and interface conditions become, using the ratio of thermal conductivities $\lambda_{sf} = \lambda_s / \lambda_f$, 
\newcommand{\Uref}{U_\text{ref}}
\newcommand{\Rref}{R_\text{ref}}
\newcommand{\Tref}{T_\text{ref}}
\begin{subequations}
	\begin{align}
		&\left. \pdv{\vartheta_s}{r}\right|_{0.5\Gamma} = - \frac{\Pe}{\lambda_{sf} \Gamma} f(\phi), \\
		&\lambda_{sf} \left. \pdv{\vartheta_s}{r}\right|_{0.5} = \left.\pdv{\vartheta_f}{r}\right|_{0.5} \quad &&\text{for all $z\in\mathbb R $ and $\phi\in[0,2\pi)$} \label{ic-nondimensional-1}, \\
		&\vartheta_f (r = 0.5,\phi,z) = \vartheta_s (r = 0.5,\phi,z) \quad&&\text{for all $z\in\mathbb R $ and $\phi\in[0,2\pi)$}.
	\end{align}
\end{subequations}
The physics of the problem in the fluid domain is governed by two dimensionless numbers, i.e. $\Re_b$ and $\Pr$. Including the solid in the formulation yields three additional dimensionless numbers: the radius ratio $\Gamma$, the ratio of thermal conductivities $\lambda_{sf}$ and a ratio of effusivities (also called thermal activity ratio, \citet{tiseljDNSTurbulentChannel2012}) $K = \sqrt{\lambda_f \rho_f c_{p,f} / \lambda_s \rho_s c_{p,s}}$. The latter describes how well temperature fluctuations can travel from the fluid to the solid and is hence only relevant in turbulent flow conditions. Note that in \cref{eq:dimensionless_temperature}, $(\rho_s c_{p,s}) / (\rho_f c_{p,f} ) = 1/(K^2 \lambda_{sf})$.

After statistical averaging, 
and the assumptions of zero mean swirl ($\avg{u_\phi} = 0$), statistically steady flow ($\pdv*{\avg{\cdot}}{t} = 0$) and fully developed conditions (no streamwise fluxes), 
\cref{eq:dimensionless_temperature} becomes: 
\begin{equation}
	0 = \frac{\lambda^*}{\Pe} \left[ \frac 1{r} \pdv{r}\left(r \pdv{\avg{\vartheta_\pp}}{r}\right) + \frac{1}{r^2} \pdv[2]{\avg{\vartheta_\pp}}{\phi}\right]  - \rho_\pp^* c_{p,\pp}^* \left[\frac{1}{r} \pdv{\left(r \avg{\vartheta_\pp{}'u_r{}'}\right)}{r} + \frac{1}{r} \pdv{\avg{\vartheta_\pp{}'u_\phi{}'}}{\phi} + 4 \avg{u_z} \overline{f}\right] 
		\label{eq:rans_temperature} \\
\end{equation}
with the second term being identically zero in the solid, which is stationary. 

The global Nusselt number $\Nu$ is defined as
\begin{align}
	\Nu &= 2 {R_i} \left. \overline{\pdv{\avg{T_f}}{r}}\right|_{R_i} \frac{1}{ T_w - T_b(z)},\label{eq:nusselt_definition}
\end{align}
where $T_b=1/(Au_b) \int_A \avg{u_z T_f} \dd A$ is the bulk temperature. Using the nondimensionalisation introduced in \cref{eq:nondim}, which is particularly useful in the fully developed state, \cref{eq:nusselt_definition} simplifies to
\begin{align}\begin{split}
		\Nu &= %
		\overline{f} \frac{\Pe}{ \vartheta_b},
	\end{split}
\end{align}
where $\vartheta_b$ is the bulk temperature non-dimensionalized according to \cref{eq:nondim}.
The (global) Nusselt number is therefore only a sensible measure if the azimuthal mean $\overline f$ of the dimensionless boundary condition $f(\phi)$ is not zero. In a similar way, one may define a local Nusselt number
\begin{align}
	\Nu_l(\phi, z) &= 2 {R_i} \left. \pdv{T_f}{r} \right|_{R_i} \frac{1}{ \left.\avg{T}\right|_{R_i} - T_{b,l}(\phi, z)},\label{eq:nusselt_local_definition}
\end{align}
where $T_{b,l}$ is the bulk temperature of a given azimuthal slice. For the case in which solid walls are not considered (no CHT), $\Nu_l$ can be directly related to $f(\phi)$:
\begin{align}
	\Nu_l(\phi, z) = - f(\phi) \frac{\Pe}{ \left.\avg{\vartheta}\right|_{0.5} - \vartheta_{b,l}(\phi)}, 
    \label{eq:nulocal_def}
\end{align}
If the solid wall is present, this is not possible since the integral solid energy equation only relates the azimuthally averaged radial heat fluxes $\left.\pdv*{\overline{\avg{T_s}}}{r}\right|_{R_i}$ and $\left.\pdv*{\overline{\avg{T_s}}}{r}\right|_{R_o}$ (note the overline). Azimuthal conduction in the solid changes the $\phi$-dependence of the heat flux between outer wall and interface.

\section{Analysis of the spectral properties of temperature moments}
\label{sec:spectral_viewpoint}
From the symmetries of the equations and the geometry, certain properties of statistical moments of temperature can be derived \textit{a priori}. In particular, these system properties are (i) linearity of temperature moments with respect to the boundary condition (due to the passive scalar assumption), (ii) translation/rotation equivariance (due to the circular cross-section), and (iii) a reflection symmetry of the equations. These properties admit a convenient treatment in wavenumber space, which simplifies both data analysis and numerical effort. Finally, $\Nu$ equivalence for boundary conditions with the same mean follows from either (i)+(ii) or (i)+(iii).

As the boundary condition and all statistics of temperature are $2\pi$-periodic in $\phi$, it is natural to employ a spectral decomposition in that direction by means of the Fourier series, as was also done in \citet{reynoldsTurbulentHeatTransfer1963} to provide a solution to the non-uniform heating problem in fully developed conditions based on the statistically averaged equations. Here, this approach is extended to understand the spectral structure of the statistical moments of temperature.

Let the complex Fourier coefficients of $f(\phi)$ be given as
\begin{equation}
	f_k = \frac{1}{2\pi} \int_{0}^{2\pi} f(\phi) \exp{-j k \phi} \mathrm{d}\phi, \label{eq:bc-fft}
\end{equation}
where $j$ denotes the imaginary unit and $k$ the azimuthal wavenumber. For a given velocity field $u_i(r, \phi, z, t)$, the temperature response is linear in the boundary condition $f(\phi)$, owing to the linearity of \eqref{eq:dimensionless_temperature}. Consequently, the full $\vartheta_\pp$ can be constructed as a linear superposition of the responses to individual Fourier modes of $f(\phi)$. First-order statistical moments in $\vartheta_\pp$ (possibly nonlinear in $u_i$), here abbreviated $\vartheta$ for readability, inherit this linearity with respect to the boundary condition. Formally, for such a moment one can write
\begin{equation}
	\avg{u_i{}^m \vartheta}(\phi) = \mathcal{L}_{\avg{u_i{}^m \vartheta}}[f(\phi) | \phi], \quad m  \in \mathbb{R}, \label{eq:linearoperator_firstordermoment}
\end{equation}
where $\mathcal{L}_{\avg{u_i{}^m \vartheta}}$ is a linear operator encoding the response of the statistic to $f(\phi)$, evaluated at angle $\phi$. Furthermore, $\mathcal{L}_{\avg{u_i{}^m \vartheta}}$ is rotation-equivariant: rotating the boundary condition by an angle $\psi$ is equivalent to rotating the moment by $\psi$:
\begin{equation}
	\mathcal{L}_{\avg{u_i{}^m \vartheta}}[f(\phi + \psi) | \phi] = \mathcal{L}_{\avg{u_i{}^m \vartheta}}[f(\phi) | \phi + \psi].
\end{equation}
A rotation-equivariant linear operator acts diagonally on the Fourier coefficients of the input 
\citep[I.8.5]{katznelsonIntroductionHarmonicAnalysis2004}. In other words, the first-order moments of temperature for a single-mode boundary condition $f(\phi) = \sin(k \phi)$ can only have a response involving the same wavenumber $k$ (possibly phase-shifted compared to $f$, e.g., for $\avg{u_\phi \vartheta}$). In particular, the azimuthal mean of all first-order moments of temperature vanishes for $k \neq 0$. Again, due to linearity, a moment that is first-order in temperature for any given composite boundary condition can be computed from the corresponding moments associated with $f(\phi) = f_k \Real{\exp{j k \phi}}$, and only the $k=0$ mode of $\phi$ can contribute to the azimuthal mean of the moment.

Therefore: (i) the Nusselt number, computed from first-order moments of temperature (see \cref{eq:nusselt_definition}), can only depend on the mean of $f(\phi)$ and not its variation over $\phi$, and (ii) if one is interested in first-order moments of temperature, instead of treating arbitrary boundary conditions $f(\phi) \in L^2[0,2\pi]$ it is sufficient to consider the response to each Fourier mode $f(\phi) = \exp{jk\phi}, k \in \mathbb Z$.

Note that all moments that appear in the Reynolds-averaged temperature equation of $\avg{\vartheta}$ are first-order in $\vartheta$, so the same linearity and Fourier-diagonal property applies (and the mean temperature equation can indeed be written in decoupled form). Likewise, transport equations for the turbulent heat flux $\avg{\vartheta'u_i}$ also do not include second-order statistics of $\vartheta$.

Importantly, this argument does \textit{not} require the governing equations to be decoupled in $\phi$, and instantaneous $u_\phi$ indeed does couple the (time-dependent) modes in turbulent flow. Turbulence is allowed to scramble azimuthal modes in a complex way; the only requirement is that the physical laws and properties governing its evolution are the same at every azimuthal position, i.e.\ that the system is equivariant under a uniform shift $\phi\mapsto\phi+\phi_0$.
These conditions - linearity and translation/rotation-equivariance, are fulfilled for a multitude of scenarios, including the thermal inlet region (which we treated previously in \cite{iTi2025}), statistically unsteady conditions, boundary layers, plane channel flows and axially rotating pipe flows. The conditions would not be met for e.g. nonzero $\pdv*{\avg{u_\phi}}{\phi}$ or $\phi$-dependent properties, such as $\phi$-dependent solid conductivity. 

For second- and higher-order moments, restrictions on the excited modes can be derived, similarly to the approach used in \cite{iTi2025}, by exploiting symmetries of the joint velocity-temperature PDF $p(u_r, u_\phi, u_z, \vartheta, r, \phi, z, t)$. For a single-mode boundary condition $f(\phi) = \sin(k \phi)$, the structure of the azimuthal spectrum of higher-order temperature statistics is constrained by symmetry considerations: only integer multiples of $k$ are excited, and parity considerations further eliminate some modes depending on the order of the moment. A derivation of this structure is given in Appendix \ref{app:higher-order}. 

Consider now a boundary condition $f(\phi) = \Real{\exp{jk\phi}}$, which is even in $\phi$. As per the discussion above, first-order moments of temperature have the same spectral shape, i.e.\ $\avg{\vartheta} = \Real{\widehat{\avg{\vartheta}}_k \exp{jk\phi}}$ and $\avg{\vartheta{}'u_i{}'} = \Real{\widehat{\avg{\vartheta{}'u_i{}'}}_k \exp{jk\phi}}$. The coefficient $\widehat{\avg{...}}_k$ is a complex-valued function of $r$ (constant in $\phi$), and encodes both gain and phase of a moment with respect to the boundary condition. If the problem has suitable additional symmetries, the phase can be explicitly determined. One such symmetry is the additional constraint of no mean rotation, i.e.\ $\avg{u_\phi} = 0$. Then, the resulting joint PDF of velocity and temperature transforms under a flip of the $\phi$ coordinate as
\begin{equation}
	p(u_r, u_\phi, u_z, \vartheta, r, \phi, z, t) = p(u_r, -u_\phi, u_z, \vartheta, r, -\phi, z, t).
\end{equation}  
Now note that
\begin{align*}
	\avg{\vartheta}(r, -\phi, z, t) &= \int_{-\infty}^{\infty} \iiint_{-\infty}^{\infty}  \vartheta p(u_r, u_\phi, u_z, \vartheta, r, - \phi, z, t) \dd u_r \dd u_\phi \dd u_z \dd \vartheta  
	\\ &= \int_{-\infty}^{\infty} \iiint_{-\infty}^{\infty}  \vartheta p(u_r, - u_\phi, u_z, \vartheta, r, \phi, z, t) \dd u_r \dd u_\phi \dd u_z \dd \vartheta 
	\\ &= \int_{-\infty}^{\infty} \iiint_{-\infty}^{\infty}  \vartheta p(u_r, u_\phi, u_z, \vartheta, r, \phi, z, t) \dd u_r \dd u_\phi \dd u_z \dd \vartheta 
	\\ &= \avg{\vartheta}(r, \phi, z, t) 
\end{align*}
Hence $\avg{\vartheta}$ is even as well, i.e.\ there is no phase shift between $f(\phi)$ and $\avg{\vartheta}$. $\avg{\vartheta{}'u_r{}'}$ and $\avg{\vartheta{}'u_z{}'}$ are also in phase with $f(\phi)$, this follows by multiplying the integrand with $u_r$ and $u_z$ in the preceding derivation. However, for $u_\phi$, one finds
\begin{align*}
	\avg{\vartheta{}u_\phi{}}(r, -\phi, z, t) &= \int_{-\infty}^{\infty} \iiint_{-\infty}^{\infty}  \vartheta u_\phi p(u_r, u_\phi, u_z, \vartheta, r, - \phi, z, t) \dd u_r \dd u_\phi \dd u_z \dd \vartheta  
	\\ &= \int_{-\infty}^{\infty} \iiint_{-\infty}^{\infty}  \vartheta u_\phi p(u_r, - u_\phi, u_z, \vartheta, r, \phi, z, t) \dd u_r \dd u_\phi \dd u_z \dd \vartheta  
	\\ &= - \int_{-\infty}^{\infty} \iiint_{-\infty}^{\infty}  \vartheta u_\phi p(u_r, u_\phi, u_z, \vartheta, r, \phi, z, t) \dd u_r \dd u_\phi \dd u_z \dd \vartheta  
	\\ &= - \avg{\vartheta{}u_\phi{}}(r, \phi, z, t)
\end{align*}
i.e.\ $\avg{\vartheta{}u_\phi{}}$ is odd; therefore $\widehat{\avg{\vartheta{}'u_\phi{}'}}_k$ must be imaginary (as $\avg{u_\phi} = 0$). Hence, the phase of $\avg{\vartheta{}'u_\phi{}'}$ is shifted by $\pi/2$ compared to the BC; for $k=0$ this component vanishes identically.

To summarise, two independent lines of argumentation have been used to reduce the spectrum of moments of temperature. The first argument establishes the first-order moments of temperature as linear and rotation-invariant operators of the boundary condition; this directly shows that the spectrum of these moments contains the same wavenumbers as the boundary condition itself, and prevents any influence of $k \neq 0$ modes in the boundary condition on the global Nusselt number. The second argument, based on the joint velocity-temperature PDF, can be used to restrict the azimuthal modes in the response of temperature moments \textit{of any order}; the resulting statement is slightly weaker for first-order moments (it still contains the statement regarding the global Nusselt number), but provides selection rules for the spectrogram of those moments for single-wavenumber boundary conditions. However, since higher-order moments are not linear in the temperature, the spectrogram of those moments for arbitrary boundary conditions containing multiple wavenumber components can be populated by non-linear interactions.  Respecting an additional constraint of no mean rotation ($\avg{u_\phi} = 0$) determines the phase of the spectrum for each first-order moment a-priori.

These considerations simplify the treatment of the azimuthally non-uniform case in \cref{sec:results} considerably. Importantly, instead of having to investigate all functions from the space of all possible boundary conditions $f(\phi)$, the first-order-in-temperature moments of \textit{any} boundary condition $f(\phi) \in L^2$ can be determined from the system response to a test boundary condition that contains the same azimuthal modes. Moreover, truncating the modal expansion after only a few dominant modes only leads to a marginal error in the bulk of the flow. The exact procedure is described in \cref{sec:cht_fdev_nonuniform}.

\section{Numerical setup}
\label{sec:numerical-setup}

\scalebox{0}{%
	\tikzsetnextfilename{pipe_legend}
	\begin{tikzpicture}
		\begin{axis}[hide axis, xmin=0, xmax=1, ymin=0, ymax=1]
			\pgfplotsinvokeforeach{IF,K0.25_L1.0,K1.0_L0.5,K1.0_L1.0,K1.0_L2.0,K4.0_L1.0,MBC,Straub_IF,Straub_MBC}{%
				\addplot[
				mark=none,
				#1, %
				] (0,0);\label{plot:#1}
			}
		\end{axis}
	\end{tikzpicture}
}

\begin{table}
	
	\begin{tabular}{lccccccc}
		\toprule
		Case      & $\Re_b$ & $L_z / R_i$ & $nel_{cs,f}$ & $nel_{cs,s}$ & $nel_z$ & $\Delta y^+_0$ & $t_{avg} u_\tau / (2R_i)$  \\ \midrule
		\verb|cht| (*) & 5300  & 100   & 132             & 72              & 216    & \SI{0.29}{}  & \SI{115.9}{} \\ 
		\verb|cht_fine| & 5300  & 25 & 384             & 240             & 95     & \SI{0.099}{} & \SI{7.0}{} \\ 
		 \bottomrule
	\end{tabular}
	\caption[Simulation metadata for pipe flow cases]{Metadata for the presented cases, including the heated streamwise length $L_z$, number of fluid (solid) elements per cross section $nel_{cs,f}$ ($nel_{cs,s}$), number of elements in the streamwise direction $nel_z$, wall distance of the first integration point in viscous units and averaging time. Data from the cases marked with (*) are used in the various figures. Extended from \cite{tsfp2026}.}
	\label{tab:pipe_cases_metadata}
\end{table}

The governing equations are solved numerically using NekRS \citep{fischerNekRSGPUacceleratedSpectral2022}, which is a portable implementation of the spectral element method. The domain dimensions, spatial resolution and averaging times are given in \cref{tab:pipe_cases_metadata}. Polynomial order $N=7$ is used with BDF2/EXT2 timestepping. The mesh of \citet{straubInfluenceThermalBoundary2019} is reused. From the perspective of the momentum equation, the present simulations are well-resolved LES simulations, where the required additional dissipation to stabilise the simulation is provided by a high-pass spectral filter as suggested by \citet{schlatterTransitionalFlowsUsing2004}. Since the thermal scales for low-$\Pr$ fluids are larger compared to the velocity scales, the thermal fields  for that case are effectively fully resolved and no subgrid model is needed for the temperature equation. This approach is called Hybrid LES/DNS or Thermal DNS by some authors \citep{straubNonhomogeneousThermalBoundary2020a}. If a solid domain is included, the azimuthal resolution is the same as in the fluid. The radial element distribution must be fine enough to resolve the decaying thermal fluctuations within the solid \citep{flageulCreationBasesDonnees2016}, and hence the first element on the solid side is mirrored from the fluid side. This choice was confirmed by means of a mesh refinement study, where the same case was run with the resolution of \citet{elkhouryDirectNumericalSimulation2013}; deviation of the mean dimensionless temperature $\avg{\vartheta}(r)$ for uniformly heated cases was found to be within \SI{0.2}{\percent} for $\Pr = 0.025$ and \SI{0.5}{\percent} for $\Pr=0.71$, while deviation of $\avg{\vartheta'\vartheta'}(r)$ was found to be within \SI{2}{\percent} for $\Pr=0.025$ and \SI{5}{\percent} for $\Pr=0.71$.\footnote{Part of this deviation may be attributed to the significantly shorter averaging time for the DNS case, see \cref{tab:pipe_cases_metadata}.} The momentum equation \cref{eq:momentum} is driven by a time-varying pressure gradient to maintain a constant bulk Reynolds number of $\Re_b = 2 u_b R_i / \nu = 5300$, corresponding to a friction Reynolds number of $\Re_\tau = u_\tau R_i / \nu \approx 180$. 

A streamwise length of $L_z = 100 R_i$ is chosen. This length is chosen as \citet{tiseljTrackingLargescaleStructures2014,straubInfluenceThermalBoundary2019} showed that for very low $\Pr$, a regular domain length of e.g.\ $L_z \approx 20R_i$ underestimates fluctuations of $\vartheta$. 
The timestep is chosen dynamically to ensure $\textit{CFL} \approx 1$. After an initial transient to the steady state, statistics are collected online. Note that convergence to the steady state can be quite slow (especially for a high thermal inertia within the solid, where no turbulent mixing occurs), so the steady state was asserted by splitting the selected averaging interval and comparing statistics.
Quantities denoted with a ${}^+$ superscript are given in viscous units; i.e.\ $\Uref = u_\tau$ and (following \cref{eq:nondim}) $\vartheta^+ = \vartheta u_\tau / u_b$. The wall distance in viscous units is defined as $y^+ = (R_i - r) / \delta_v$ where $\delta_v$ is the viscous length scale, i.e.\ negative values of $y^+$  denote points inside the solid.

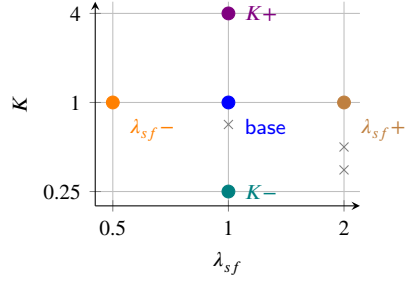
\begin{figure}
	\centering
  \begin{tikzpicture}
	\begin{axis}[
		width=3.5cm,
		height=2.7cm,
		scale only axis,
		clip mode=individual,
		grid=both,
		xmode=log,
		ymode=log,
		xmin=0.45,
		xmax=2.20,
		ymin=0.20,
		ymax=4.80,
		xlabel={$\lambda_{sf}$},
		ylabel={$K$},
		xtick={0.5,1,2},
		log ticks with fixed point,
		ytick={0.25,1,4},
		minor tick num=0,
		axis lines=left,
		every axis plot/.append style={only marks, mark=*, mark size=2.4pt},
		]
        \addplot[only marks, mark=x, black!50] coordinates {
      (1, 1)
      (2, 0.5)
      (1, 0.71)
      (2, 0.35)
    };\label{plot:cruzpoints}
		\addplot[K1.0_L0.5] coordinates {(0.5,1)};
		\addplot[K1.0_L2.0] coordinates {(2,1)};
		\addplot[K1.0_L1.0] coordinates {(1,1)};
		\node[below right=0.1cm, text=K1.0_L0.5color] at (axis cs:0.5,1) {$\lambda_{sf}-$};
		\node[below right=0.1cm, text=K1.0_L2.0color] at (axis cs:2,1) {$\lambda_{sf}+$};
		\node[below right=0.1cm, text=K1.0_L1.0color] at (axis cs:1,1) {base};

			\addplot[K0.25_L1.0] coordinates {(1,0.25)};
			\addplot[K4.0_L1.0] coordinates {(1,4)};
			\node[right=0.1cm, text=K0.25_L1.0color] at (axis cs:1,0.25) {$K-$};
			\node[right=0.1cm, text=K4.0_L1.0color] at (axis cs:1,4) {$K+$};

	\end{axis}
\end{tikzpicture}%
\caption{Considered cases for conjugate heat transfer, parametrised by $K$ and $\lambda_{sf}$. 
The parameter combinations marked with \pgfplotsplotfromname {plot:cruzpoints} were considered by \cite{vicentecruzVersatileImmersedBoundary2023}.}
\label{tab:pipe_cases_linestyles}
\end{figure}

A radius ratio of $\Gamma = 1.2$ is used throughout as in \citet{vicentecruzVersatileImmersedBoundary2023};  a similar value of $\Gamma = 1.225$ is found in the experimental facility GALINKA at KIT \citep{laubeTurbulentHeatTransfer2022}. The considered CHT cases, i.e.\ representative combinations of solid properties, are summarized in \cref{tab:pipe_cases_linestyles}, where the configurations investigated by \citet{vicentecruzVersatileImmersedBoundary2023} are included in gray for reference. We start with a baseline configuration where fluid and solid properties are equal ($\lambda_{sf} = K = 1$). The conductivity ratio $\lambda_{sf}$ is then varied as $\{0.5, 2\}$ for constant $K = 1$, and the thermal activity ratio $K$ is varied as $\{0.25, 4\}$ for constant $\lambda_{sf} = 1$.\footnote{Typical values for air at ambient conditions ($\Pr \approx 0.71$) are $\lambda_{sf} \approx \SI{1.6e4}{}\ (\SI{3.6e3}{})$, $K \approx  \SI{1.5e-4}{}\ (\SI{3.0e-4}{})$ for a Copper (Nickel) wall; for water ($\Pr \approx 7$): $\lambda_{sf} \approx \SI{670}{}\ (\SI{157}{})$, $K \approx  \SI{0.043}{}\ (\SI{0.083}{})$ and for GaInSn ($\Pr \approx 0.32$): $\lambda_{sf} \approx \SI{17}{}\ (\SI{3.9}{})$, $K \approx  \SI{0.20}{}\ (\SI{0.40}{})$. Values of $\lambda_{sf} < 1$ and $K > 1$ are obtained for e.g GaInSn paired with steel 1.4828: $\lambda_{sf} \approx 0.49, K \approx 1.1$. Material constants taken from  \citet{stephanVDIWaermeatlasFachlicherTraeger2019,LaubeDietrichMarocco2023_1000155030}.} 
In addition, a case without the solid, i.e.\ the heat flux is prescribed on the fluid directly (isoflux -- IF), is considered for the various boundary conditions. This corresponds to a very thin solid through which temperature fluctuations can pass through unhindered (very low thermal inertia). For the mixed boundary condition (MBC) case, the same forcing of the temperature equation is applied, prescribing the same mean heat flux, but the wall temperature is kept constant at a uniform value. This corresponds to the limiting case of a very thin solid with very high thermal inertia, which completely blocks temperature fluctuations from passing into the solid and hence from existing at the wall at all, and it is only considered for the case of uniform heat flux.

\section{Numerical results and discussion}
\label{sec:results}

This section presents well-resolved numerical results for conjugate heat transfer under fully developed conditions, with azimuthally uniform and non-uniform heat flux. As described above, \cref{tab:pipe_cases_linestyles} summarizes the investigated cases in terms of the dimensionless quantities $\lambda_{sf}$ and $K$, introduced by the presence of the solid. The radius ratio $\Gamma$ is kept constant at $\Gamma = 1.2$. In the following, the colours introduced in \cref{tab:pipe_cases_linestyles} are used to distinguish the different cases in the plots in \cref{sec:uniformhf,sec:cht_fdev_nonuniform}. Hence the base case  with $K=1$ and $\lambda_{sf} = 1$ (i.e., equal solid and fluid material properties) is shown in blue, while a variation of the solid conductivity ratio (at fixed $K=1$) to $\lambda_{sf} = 2$ is visualized in brown and to $\lambda_{sf} = 0.5$ in orange. A variation of the effusivity ratio (at $\lambda_{sf} = 1$) to $K=4$ is plotted in purple and to $K=0.25$ in green. Additionally, the results include cases without conjugate heat transfer as a reference:  prescribed fluid heat flux (IF, dashed red) and mixed boundary condition (MBC, dashed light green) where the latter is only included for uniform heat flux.

\subsection{Uniform heat flux}
\label{sec:uniformhf}

This subsection considers $f(\phi) =1$ under fully developed conditions. 
This configuration was previously studied in \citet{vicentecruzVersatileImmersedBoundary2023} for $\lambda_{sf} \in \{1, 2\}$ and $1/(K^2 \lambda_{sf}) \in \{1, 2\}$ using an immersed boundary method implemented in XCompact3D. Here, we consider an expansion of the parameter space, particularly towards $K > 1$ and $\lambda_{sf} < 1$ (see \cref{tab:pipe_cases_linestyles} for a comparison of the parameter spaces, where the configurations of \citet{vicentecruzVersatileImmersedBoundary2023} are shown in gray). 
\newcommand{\plotHomTempProfileAll}[1]{%

      \pgfplotstableread[col sep=comma]{#1}\datatable
      \pgfplotsinvokeforeach{IF,K0.25_L1.0,K1.0_L0.5,K1.0_L2.0,K4.0_L1.0,K1.0_L1.0,MBC}{%
        \addplot[
          mark=none,
          ##1, %
        ] table[x={y+}, y={##1_0}] from \datatable;
      }
}
\newcommand{\plotHomTempProfileAllSeparate}[1]{%
  \pgfplotstableread[col sep=comma]{#1}\datatable
      \pgfplotsinvokeforeach{K0.25_L1.0,K1.0_L0.5,K1.0_L2.0,K1.0_L1.0,K4.0_L1.0}{%
        \addplot[
          mark=none, smooth,
          ##1, %
          restrict x to domain=-1000:0,
        ] table[x={y+}, y={##1_0}] from \datatable;
      }
      \pgfplotsinvokeforeach{IF,K0.25_L1.0,K1.0_L0.5,K1.0_L2.0,K4.0_L1.0,K1.0_L1.0,MBC}{%
        \addplot[
          mark=none, smooth,
          ##1, %
          restrict x to domain=0:1000,
        ] table[x={y+}, y={##1_0}] from \datatable;
      }
}

\newcommand{\plotHomTempProfileStraub}[1]{%

      \pgfplotstableread[col sep=comma]{#1}\datatable
      \pgfplotsinvokeforeach{IF,MBC}{%
        \addplot[
          Straub_##1, %
        ] table[x={y+}, y={##1}] from \datatable;
      }
}

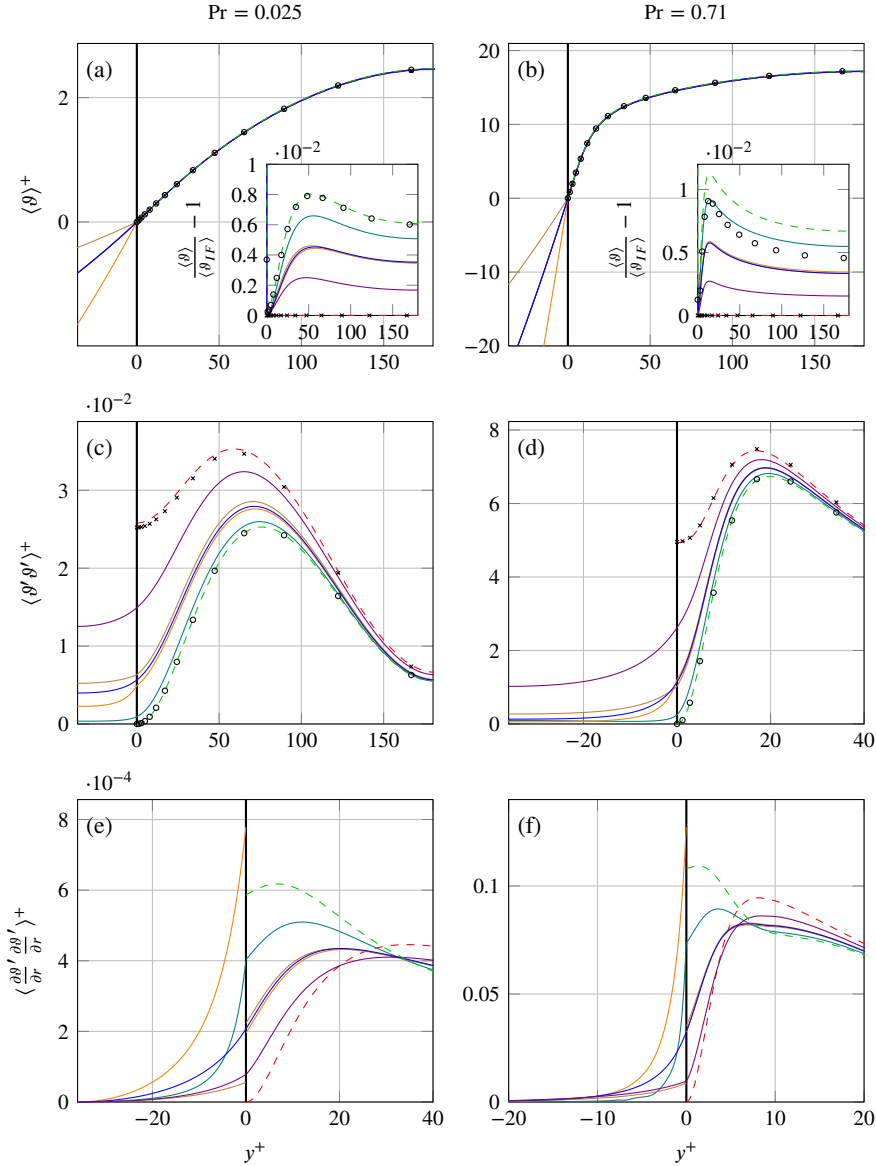
\begin{figure}
\centering
\tikzsetnextfilename{onept_stats_hom}
    \begin{tikzpicture}
        \begin{groupplot}[group style={group size=2 by 3, group name=my plots}, width=4.7cm,
      height=4cm,
      xmin=-36,
      xmax=180,
      scale only axis,
      grid=both,
      ]
            \nextgroupplot[ylabel={$\avg{\vartheta}^+$},clip=false, title={$\Pr = 0.025$}, subplotlabel=fig:meantemp_pr0025]
            \draw[thick] (0, \pgfkeysvalueof{/pgfplots/ymin}) -- (0, \pgfkeysvalueof{/pgfplots/ymax}); %
            \plotHomTempProfileAll{figdata/pipe/retau_180_les_r1.2_long/hom_temp_profile_Pr_0.025.csv}
            \plotHomTempProfileStraub{figdata/pipe/retau_180_les_r1.2_long/straub_temp_profile_Pr_0.025.csv}
            \nextgroupplot[clip=true, ymin=-20, title={$\Pr = 0.71$}, subplotlabel=fig:meantemp_pr071]
            \draw[thick] (0, \pgfkeysvalueof{/pgfplots/ymin}) -- (0, \pgfkeysvalueof{/pgfplots/ymax}) node[above right] {fluid} node[above left] {solid};
            \plotHomTempProfileAll{figdata/pipe/retau_180_les_r1.2_long/hom_temp_profile_Pr_0.71.csv}
            \plotHomTempProfileStraub{figdata/pipe/retau_180_les_r1.2_long/straub_temp_profile_Pr_0.71.csv}
            \nextgroupplot[clip=true, ylabel={$\avg{\vartheta'\vartheta'}^+$}, ymin=0, subplotlabel=fig:trms_pr0025]
            \draw[thick] (0, \pgfkeysvalueof{/pgfplots/ymin}) -- (0, \pgfkeysvalueof{/pgfplots/ymax}) node[above right] {fluid} node[above left] {solid};
            \plotHomTempProfileAll{figdata/pipe/retau_180_les_r1.2_long/hom_trms_profile_Pr_0.025.csv}
            \plotHomTempProfileStraub{figdata/pipe/retau_180_les_r1.2_long/straub_trms_profile_Pr_0.025.csv}
            \nextgroupplot[clip=true, ymin=0, xmax=40, subplotlabel=fig:trms_pr071]
            \draw[thick] (0, \pgfkeysvalueof{/pgfplots/ymin}) -- (0, \pgfkeysvalueof{/pgfplots/ymax}) node[above right] {fluid} node[above left] {solid};
            \plotHomTempProfileAll{figdata/pipe/retau_180_les_r1.2_long/hom_trms_profile_Pr_0.71.csv}
            \plotHomTempProfileStraub{figdata/pipe/retau_180_les_r1.2_long/straub_trms_profile_Pr_0.71.csv}
            \nextgroupplot[ylabel={$\avg{\pdv{\vartheta}{r}'\pdv{\vartheta}{r}'}^+$}, xlabel={$y^+$}, clip=true, ymin=0, xmin=-36, xmax=40, subplotlabel=fig:dtdrrms_pr0025]
            \draw[thick] (0, \pgfkeysvalueof{/pgfplots/ymin}) -- (0, \pgfkeysvalueof{/pgfplots/ymax}) node[above right] {fluid} node[above left] {solid};
            \plotHomTempProfileAllSeparate{figdata/pipe/retau_180_les_r1.2_long/hom_dtdr_rms_profile_Pr_0.025.csv}
            \nextgroupplot[clip=true, ymin=0,  xlabel={$y^+$}, ymin=0, xmin=-20, xmax=20, subplotlabel=fig:dtdrrms_pr071, yticklabel style={
			        /pgf/number format/fixed,
			        /pgf/number format/precision=2
			},]
            \draw[thick] (0, \pgfkeysvalueof{/pgfplots/ymin}) -- (0, \pgfkeysvalueof{/pgfplots/ymax}) node[above right] {fluid} node[above left] {solid};
            \plotHomTempProfileAllSeparate{figdata/pipe/retau_180_les_r1.2_long/hom_dtdr_rms_profile_Pr_0.71.csv}
        \end{groupplot}
        \begin{axis}[anchor=south east, at={($(my plots c1r1.south east)+(-0.2cm,0.4cm)$)}, ylabel={$\frac{\avg{\vartheta}}{\avg{\vartheta_{IF}}} - 1$}, 
        axis background/.style={fill=white}, width=2cm, ymin=0,
        xmin=0, xmax=180, height=2cm, scale only axis, ymax=0.01]
            \plotHomTempProfileAll{figdata/pipe/retau_180_les_r1.2_long/hom_temp_diff_profile_Pr_0.025.csv}
              \plotHomTempProfileStraub{figdata/pipe/retau_180_les_r1.2_long/straub_temp_diff_profile_Pr_0.025.csv}
        \end{axis}
        \begin{axis}[anchor=south east, 
        at={($(my plots c2r1.south east)+(-0.2cm,0.4cm)$)}, ylabel={$\frac{\avg{\vartheta}}{\avg{\vartheta_{IF}}} - 1$},
        axis background/.style={fill=white}, width=2cm, ymin=0,
        xmin=0, xmax=180, height=2cm, scale only axis, ymax=0.012]
            \plotHomTempProfileAll
              {figdata/pipe/retau_180_les_r1.2_long/hom_temp_diff_profile_Pr_0.71.csv}
              \plotHomTempProfileStraub
              {figdata/pipe/retau_180_les_r1.2_long/straub_temp_diff_profile_Pr_0.71.csv}
        \end{axis}
    \end{tikzpicture}
    \caption[Radial profiles of $\avg{\vartheta}^+$, $\avg{\vartheta'\vartheta'}^+$, and $\avg{\pdv*{\vartheta'}{r}\pdv*{\vartheta'}{r}}$ for $\Re_b = 5300$, varying $\Pr$ and different solid properties.]{Mean profiles of various quantities for homogeneous heat flux at $\Re_b = 5300$ (LES); left: $\Pr = 0.025$, right: $\Pr = 0.71$. First row: Mean profiles of dimensionless temperature and (inset) comparison with IF as baseline. Second row: temperature variance, third row: variance of $\pdv*{\vartheta}{r}$. Colors indicate varying solid properties as given in \cref{tab:pipe_cases_linestyles}. Additionally, data from \citet{straubInfluenceThermalBoundary2019} is included where available (DNS, $L_z = 12.5D$): IF (\pgfplotsplotfromname{plot:Straub_IF}) and MBC (\pgfplotsplotfromname{plot:Straub_MBC}).}
\end{figure}
Results in terms of  the dimensionless mean temperature in inner units, for the various cases defined in \cref{tab:pipe_cases_linestyles}, are presented in \cref{fig:meantemp_pr0025,fig:meantemp_pr071} for two different Prandtl numbers. Recall that negative values of $y^+$ correspond to the solid region. The temperature distribution in the solid is known a priori (linear conduction solution based on an integration of \cref{eq:dimensionless_temperature} over $z$ and $\phi$ which shows a dependence on   $\lambda_{sf}$ but not on $K$). In consequence, data for varying $K$ (blue, green, purple) at $\lambda_{sf}=1$ collapse within the solid. 

For the fluid side, \citet{straubInfluenceThermalBoundary2019} observed that the mean temperature profiles are very similar for IF and MBC cases. This data is included for reference. The existing, but small, dependence of the mean temperature field on the solid properties are visualized in the insets of  \cref{fig:meantemp_pr0025,fig:meantemp_pr071} in terms of a deviation from the IF case. As expected, the present DNS data for varying $K$ are located in between IF and MBC. 
There is virtually no dependence on $\lambda_{sf}$, while a small dependence on $K$ is visible.  The very small sensitivity of $\vartheta$ to $\lambda_{sf}$ and $K$ implies that the heat transfer in terms of $\Nu$ does not carry a significant dependence on the solid's properties in fully developed pipe flows.

The temperature variance $\avg{\vartheta'\vartheta'}^+$, however, is strongly affected by the solid's properties: while MBC (and $K \rightarrow 0$) removes fluctuations at the wall, they can travel freely between fluid and solid for IF, and indeed an increase of $K$ increases the level of fluctuations in both subdomains (solid and fluid). This quantity is displayed in \cref{fig:trms_pr0025,fig:trms_pr071}.\footnote{Note that in \cref{fig:trms_pr0025}, i.e.\ $\avg{\vartheta'\vartheta'}$ for $\Pr=0.025$, the DNS data by \citet{straubInfluenceThermalBoundary2019} is slightly offset from the present data for the IF boundary condition. We attribute this offset to the shorter domain length used in the reference DNS, not to grid resolution.} At least in the fluid domain, there is little influence of $\lambda_{sf}$ on $\avg{\vartheta'\vartheta'}^+$. For  low-$\Pr$ fluids temperature fluctuations reach farther into the solid for the same $\{K, \lambda_{sf}\}$.  Similar trends can be seen for the variance of the temperature gradient $\avg{(\pdv*{\vartheta}{r})'{}^2}$, displayed in \cref{fig:dtdrrms_pr0025,fig:dtdrrms_pr071}. At the wall, this quantity is zero for IF (as  $\pdv*{\vartheta}{r}$ is fixed in time by the boundary condition) and maximum for MBC. If $\lambda_{sf} \neq 1$, this quantity exhibits a jump at the wall, as is expected from \cref{ic-nondimensional-1} which directly translates to an interface condition for the fluctuation of $\pdv*{\vartheta}{r}$. These results are in agreement with those of \citet{vicentecruzVersatileImmersedBoundary2023} and \citet{straubInfluenceThermalBoundary2019}.

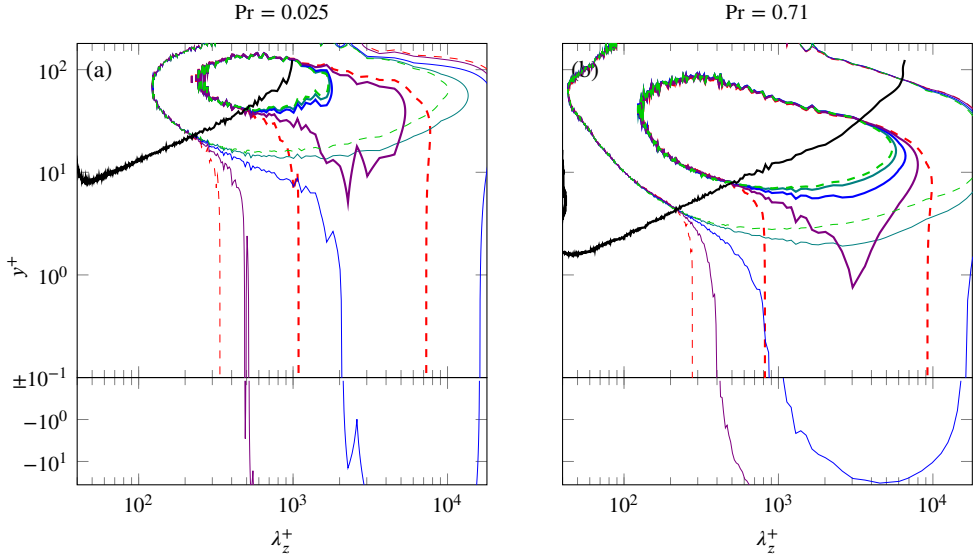
\begin{figure}
\centering
\tikzsetnextfilename{spectra_hom}
\begin{tikzpicture}

		\def\xmin{40}
		\def\xmax{18000}

		\begin{axis}[
			y filter/.expression={ln(-y)},
			y dir=reverse,
			name=low,
			xlabel={$\lambda_z^+$},
			xmin=\xmin, xmax=\xmax,
			xmode=log,
			ymode=log,
			ymin=0.1, ymax=36,           %
			ytick={0.1,1,10,100},
			yticklabels={,$-10^0$,$-10^1$,$-10^2$},
			width=7cm, height=3cm,
			enlargelimits=false,
			clip=true,
			point meta min=0,
			point meta max=1,
			contour level style on,
            restrict y to domain=-2:5,
			log basis y=10,
			]
			
			\foreach \name/\dash in {
				IF/dashed,
				K0.25_L1.0/solid,
				K1.0_L1.0/solid,
				K4.0_L1.0/solid,
				MBC/dashed
			}{%
				\edef\temp{%
					\noexpand\addplot[
					contour prepared={labels=false, draw color=\name color},
					\dash
					] table {figdata/pipe/retau_180_les_r1.2_long/t_zspectra_les_r1.2_0.025_\name_0.pgf};%
				}%
				\temp
			}
		\end{axis}
		\begin{axis}[
			name=up,
			title={$\Pr=0.025$}, subplotlabel=spectra_pr0025,
			at={(low.north west)}, anchor=south west,
			xmin=\xmin, xmax=\xmax,
			xmode=log,
			ymode=log,
			ymin=0.1, ymax=180,           %
			ytick={0.1, 1,10,100},
			yticklabels={$\pm 10^{-1}$,$10^0$,$10^1$,$10^2$},
			xticklabels=\empty,
			width=7cm, height=6cm,
			enlargelimits=false,
			clip=true,
			point meta min=0,
			point meta max=1,
			y label style={at={(-0.1,0.333)}},
			ylabel={$y^+$},
			contour level style on,
            restrict y to domain=-2:5,
			log basis y=10,
			]
			\foreach \name/\dash in {
				IF/dashed,
				K0.25_L1.0/solid,
				K1.0_L1.0/solid,
				K4.0_L1.0/solid,
				MBC/dashed
			}{%
				\edef\temp{%
					\noexpand\addplot[
					contour prepared={labels=false, draw color=\name color},
					\dash
					] table {figdata/pipe/retau_180_les_r1.2_long/t_zspectra_les_r1.2_0.025_\name_0.pgf};%
				}%
				\temp
			}
            \addplot[contour prepared={labels=false, draw color=black}] table {figdata/pipe/retau_180_les_r1.2_long/t_spectra_const_0.025.pgf};
			
		\end{axis}

		\begin{axis}[
			y filter/.expression={ln(-y)},
			y dir=reverse,
            at={($(low.north east) + (1cm, 0)$)}, 
            anchor=north west,
			name=low2,xlabel={$\lambda_z^+$},
			xmin=\xmin, xmax=\xmax,
			xmode=log,
			ymode=log,
			ymin=0.1, ymax=36,           %
			ytick={0.1,1,10,100},
			yticklabels=\empty, %
			width=7cm, height=3cm,
			enlargelimits=false,
			clip=true,
			point meta min=0,
			point meta max=1,
            contour level style on,
            restrict y to domain=-2:5,
			log basis y=10,
			]
			
			\foreach \name/\dash in {
				IF/dashed,
				K0.25_L1.0/solid,
				K1.0_L1.0/solid,
				K4.0_L1.0/solid,
				MBC/dashed
			}{%
				\edef\temp{%
					\noexpand\addplot[
					contour prepared={labels=false, draw color=\name color},
					\dash
					] table {figdata/pipe/retau_180_les_r1.2_long/t_zspectra_les_r1.2_0.71_\name_0.pgf};%
				}%
				\temp
			}

		\end{axis}

		\begin{axis}[
			name=up,
			at={(low2.north west)}, anchor=south west,
			title={$\Pr=0.71$}, subplotlabel=spectra_pr071,
			xmin=\xmin, xmax=\xmax,
			xmode=log,
			ymode=log,
			ymin=0.1, ymax=180,           %
			ytick={0.1, 1,10,100},
			yticklabels=\empty,%
			xticklabels=\empty,
			width=7cm, height=6cm,
			enlargelimits=false,
			clip=true,
			point meta min=0,
			point meta max=1,
            contour level style on,
            restrict y to domain=-2:5,
			log basis y=10,
			]
			\foreach \name/\dash in {
				IF/dashed,
				K0.25_L1.0/solid,
				K1.0_L1.0/solid,
				K4.0_L1.0/solid,
				MBC/dashed
			}{%
				\edef\temp{%
					\noexpand\addplot[
					contour prepared={labels=false, draw color=\name color},
					\dash
					] table {figdata/pipe/retau_180_les_r1.2_long/t_zspectra_les_r1.2_0.71_\name_0.pgf};%
				}%
				\temp
			}
            \addplot[contour prepared={labels=false, draw color=black}] table {figdata/pipe/retau_180_les_r1.2_long/t_spectra_const_0.71.pgf};
			
		\end{axis}
		
	\end{tikzpicture}

\caption[Streamwise spectra of the temperature field for $\Re_b = 5300$, varying $\Pr$ and different solid properties.]{Streamwise spectra of the temperature field. (a): $\Pr = 0.025$,   (b): $\Pr=0.71$. Spectra are computed from at least 30 independent snapshots. The figures display the cases of varying $K$ as well as the limiting cases of MBC and IF, where the line styles are given in \cref{tab:pipe_cases_linestyles}. Contour levels at 0.1 (thin) / 0.5 (thick lines) of the individual maximum are displayed. Note that in the top left of the individual diagrams, the spectral energies of the respective cases are in agreement with each other; the boundary of this region is indicated by the black solid line (5\% deviation between IF and MBC).}
\label{fig:spectra_hom}
\end{figure}

$\avg{\vartheta'\vartheta'}$ is further analysed by decomposing it into its axial spectrum (\cref{fig:spectra_hom}). Streamwise power spectral densities of the temperature field are computed as the product of the streamwise Fourier coefficients of the instantaneous temperature with their complex conjugate (i.e.\ $E = \hat \vartheta \hat \vartheta^*$) for both $\Pr$ and varying solid properties.  The quantity displayed is the premultiplied spectral energy density for the cases IF, MBC and those with $\lambda_{sf}= 1$ (i.e.\ varying $K$) for both $\Pr$ numbers. \Cref{fig:spectra_hom} displays contour lines of the premultiplied power spectral densities (i.e.\ $k_z E$) at $\SI{10}{\percent}$ and $\SI{50}{\percent}$ of the maximum value; the horizontal axis is the inner-scaled axial wavelength $\lambda_z^+ = 2 \pi / k_z$. 

With increasing wall distance, the energy contained in fluctuations of increasingly large axial wavelength becomes independent of $K$ for both $\Pr$. At $\Pr = 0.71$, $K$ has no impact on the spectral energy density in the pipe's core for fluctuations smaller than $\lambda_z^+ \lesssim 6000$ ($\lambda_z / R_i \lesssim 33$). Conceptually, one can imagine the temperature spectrum at a given (sufficiently large) wall distance $\tilde y^+$ as the near-wall spectrum filtered by the accumulated turbulent mixing up to $\tilde y^+$; making all but the largest axial fluctuations indistinguishable from an MBC reference. For $\Pr = 0.025$, the same effect is observed; but the enhanced conduction reduces the effectiveness of the filter: near-wall, instantaneous temperature information can travel further towards the pipe's core by means of conduction. In this case, the spectral energy density in the pipe's core becomes independent of $K$ for $\lambda_z^+ \lesssim 1000$ ($\lambda_z / R_i \lesssim 5.5$). Consequently, there is stronger effect of the solid's properties on the temperature fluctuations in the pipe's core for $\Pr=0.025$, as seen in the overall profiles (\cref{fig:trms_pr0025,fig:trms_pr071}).

\subsection{Non-uniform heat flux}
\label{sec:cht_fdev_nonuniform}
In this section, azimuthally non-uniform heat flux is considered. 
As per \cref{sec:spectral_viewpoint}, the azimuthal mean of statistics that are first-order in $\vartheta$ -- such as the azimuthally averaged mean temperature $\overline{\avg{\vartheta}}$ -- are independent of the distribution of the heat flux around the circumference. In consequence, the results of \cref{sec:uniformhf} apply with respect to $\overline{\avg{\vartheta}}$ for any solid properties. However, any $\phi$-resolved statistical moment does depend on $f(\phi)$, and, of course, on $\Re_b$, $\Pr$ and the dimensionless properties introduced by the presence of the solid, i.e. $\Gamma$, $\lambda_{sf}$ and $K$.

Based on the Fourier-diagonality property of $\mathcal{L}_{\avg{u_i^m \vartheta}}$ discussed in \cref{sec:spectral_viewpoint}, two key simplifications can be made to investigate $f(\phi)$ effects. First, by determining the system response to a composite boundary condition, one can obtain all modal responses (of first-order temperature statistics) of the contained wavenumbers by means of a Fourier transform. For simplicity, this test boundary condition is chosen as $f(\phi) = \sum_{k=1}^{10} \sin(k \phi)$, and from this one obtains the (modal) system response for $\avg{\widehat{u_i^m \vartheta}}_k$, $1 \leq k \leq 10$. This is conceptually similar to characterizing a linear, time-invariant dynamical system by its impulse response (that contains all frequencies). Second, one can assemble $\avg{u_i^m \vartheta}(\phi)$ for any arbitrary boundary condition $f(\phi)$ from a Fourier-coefficient-weighted sum of the modal responses $\avg{\widehat{u_i^m \vartheta}}_k$.\footnote{As a validation step, the expected response to the half-sinusoidal boundary condition 
\begin{equation*}
	f(\phi) = \begin{cases}
		\pi \sin(\phi) & \phi \in [0, \pi], \\ 
		0 & \text{otherwise},
	\end{cases} \label{eq:halfsinbc}
\end{equation*}
is computed based on the response to the test boundary condition and the response to uniform heating, and compared to the actual response to the half-sinusoidal boundary condition. For $\avg{\vartheta}$, the root mean square of the deviation is less than $\SI{0.5}{\percent}$ of the centreline temperature. The main cause of the deviation is the omission of wavenumbers $k > 10$ which is only relevant near the wall as these decay fast with increasing $y^+ > 0$, see discussion of \cref{fig:mean_temp_modes} below (in \cite{neuhauserSolutionExtendedGraetz2025}, the same was observed for laminar flow). }

In this subsection, these modal responses $\avg{\widehat{u_i^m \vartheta}}_k$ are presented for the first few, and usually dominant, wavenumbers, and the impact of solid and fluid properties on those is characterised. The full dataset \cite{neuhauserDatasetConjugateHeat2026} contains complete first-order statistics for $0 \leq k \leq 10$.

\newcommand{\plotHomTempProfileAllK}[3]{%
	\pgfplotstableread[col sep=comma]{#1}\datatable
	\pgfplotsinvokeforeach{IF,K0.25_L1.0,K1.0_L0.5,K1.0_L1.0,K1.0_L2.0,K4.0_L1.0}{%
		\addplot[
		##1,
		#3,
		] table[x={y+}, y={##1_5_k#2}] from \datatable;
	}
}

\begin{figure}
	\centering
	\tikzsetnextfilename{azimuthal_modes_turbulent}

\begin{tikzpicture}
	\begin{groupplot}[group style={group size=2 by 2, horizontal sep=0.8cm}, xlabel=$y^+$, xmin=-36, xmax=181, width=5.8cm, height=7cm, title={$\Pr=0.025$}, subplotlabel=fig:mean_t_kcomp_0025] 
	\nextgroupplot[ ylabel=$\avg{\hat{\vartheta}}_k^+$]
		\plotHomTempProfileAllK{figdata/pipe/retau_180_les_r1.2_long/decomp_t_Pr0.025.csv}{0}{}
		\plotHomTempProfileAllK{figdata/pipe/retau_180_les_r1.2_long/decomp_t_Pr0.025.csv}{1}{}
		\plotHomTempProfileAllK{figdata/pipe/retau_180_les_r1.2_long/decomp_t_Pr0.025.csv}{2}{}
		\plotHomTempProfileAllK{figdata/pipe/retau_180_les_r1.2_long/decomp_t_Pr0.025.csv}{6}{}
		\draw (50, -2.4) -- (60, -3) node[right ] {$k=1$};
		\draw (50, -0.75) -- (60, -1.25) node[right] {$k=2$};
		\draw (50, 0) -- (60, 0.5) node[right] {$k=6$};
		\draw (50, 1.2) -- (60, 2.1) node[right] {$k=0$};
        \draw[thick] (0, \pgfkeysvalueof{/pgfplots/ymin}) -- (0, \pgfkeysvalueof{/pgfplots/ymax}) node[above right] {fluid} node[above left] {solid};
	\nextgroupplot[ymin=-30, title={$\Pr=0.71$}, subplotlabel=fig:mean_t_kcomp_071, name=g2]	
		\plotHomTempProfileAllK{figdata/pipe/retau_180_les_r1.2_long/decomp_t_Pr0.71.csv}{0}{}
		\plotHomTempProfileAllK{figdata/pipe/retau_180_les_r1.2_long/decomp_t_Pr0.71.csv}{1}{}
		\plotHomTempProfileAllK{figdata/pipe/retau_180_les_r1.2_long/decomp_t_Pr0.71.csv}{2}{}
		\plotHomTempProfileAllK{figdata/pipe/retau_180_les_r1.2_long/decomp_t_Pr0.71.csv}{6}{}

		\draw (50, -7) -- (60, -8) node[right ] {$k=1$};
		\draw (50, -2.5) -- (60, -3) node[right] {$k=2$};
		\draw (50, 0) -- (60, 3) node[right] {$k=6$};
		\draw (50, 13.5) -- (60, 11.5) node[right] {$k=0$};
        \draw[thick] (0, \pgfkeysvalueof{/pgfplots/ymin}) -- (0, \pgfkeysvalueof{/pgfplots/ymax}) node[above right] {fluid} node[above left] {solid};
        \end{groupplot}

    \node[anchor=north west, name=k0] at ($(g2.north east) + (0.5, -0.4)$)  {
		\begin{tikzpicture}[scale=0.8]
        \draw[fill=black!20] (0, 0) circle (1);
        \draw[fill=white] (0,0) circle ({1/1.2});
			\draw[thick, use as bounding box] (0, 0) circle (1);
			\draw[red] plot[domain=0:360, samples=50] (\x:{1*0.3+1});
			\foreach \phi in {15, 30, ..., 360}{
				\draw[->, red] (\phi:{1*0.3+1}) -- (\phi:1); }
			\draw[dashed] (0, 0) -- (1, 0);
			\draw[fill] (0,0) circle (0.05);
			\draw (0, 1.4) node[above] {$k=0$};
		\end{tikzpicture}};
	
	\node[anchor=north west, name=k1] at ($(k0.north east) + (0.5, 0)$) {
		\begin{tikzpicture}[scale=0.8]
        \draw[fill=black!20] (0, 0) circle (1);
        \draw[fill=white] (0,0) circle ({1/1.2});
			 \draw[thick, use as bounding box] (0, 0) circle (1);
			\draw[red] plot[domain=0:360, samples=50] (\x:{cos(\x)*0.3+1});
			\foreach \phi in {30, 45, ..., 150, 210, 225, ..., 330}{
				\draw[->, red] (\phi-90:{sin(\phi)*0.3+1}) -- (\phi-90:1); }
			\draw[dashed] (0, 0) -- (1, 0);
			\draw[fill] (0,0) circle (0.05);
			\draw (0, 1.4) node[above] {$k=1$};
	\end{tikzpicture}};

	\node[anchor=north west] at ($(k0.south west) + (0, -1.1)$) {
		\begin{tikzpicture}[scale=0.8]
        \draw[fill=black!20] (0, 0) circle (1);
        \draw[fill=white] (0,0) circle ({1/1.2});
			\draw[thick, use as bounding box] (0, 0) circle (1);
			\draw[red] plot[domain=0:360, samples=50] (\x:{cos(2*\x)*0.3+1});
			\foreach \phi in {15, 30, ..., 75, 105, 120, ..., 165, 195, 210, ..., 255, 285, 300, ..., 345}{
				\draw[->, red] (\phi-45:{sin(2*\phi)*0.3+1}) -- (\phi-45:1); }
			\draw[dashed] (0, 0) -- (1, 0);
			\draw[fill] (0,0) circle (0.05);
			\draw (0, 1.3) node[above] {$k=2$};
	\end{tikzpicture}};

	\node[anchor=north west] at ($(k1.south west) + (0, -1.1)$) {
		\begin{tikzpicture}[scale=0.8]
        \draw[fill=black!20] (0, 0) circle (1);
        \draw[fill=white] (0,0) circle ({1/1.2});
			\draw[thick, use as bounding box] (0, 0) circle (1);
			\draw[red] plot[domain=0:360, samples=200] (\x:{cos(6*\x)*0.3+1});
			\foreach \phi in {15, 45, ..., 345}{
				\draw[->, red] ({\phi-15}:{sin(6*\phi)*0.3+1}) -- (\phi-15:1); }
			\draw[dashed] (0, 0) -- (1, 0);
			\draw[fill] (0,0) circle (0.05);            
			\draw (0, 1.3) node[above] {$k=6$};
        \end{tikzpicture}};
\end{tikzpicture}
	
	\caption{Selected azimuthal modes of the mean temperature operator at $\Re_b = 5300$. (a): $\Pr = 0.025$, (b): $\Pr = 0.71$. Line styles/colors correspond to different solid properties; see \cref{tab:pipe_cases_linestyles}. Right: visualisation of the boundary conditions corresponding to $k \in \{0, 1, 2, 6\}$; the displayed modal temperature $\avg{\hat \vartheta}_k$ can be understood as radial profile at the azimuthal location marked with the dashed line. }
	\label{fig:mean_temp_modes}
\end{figure}
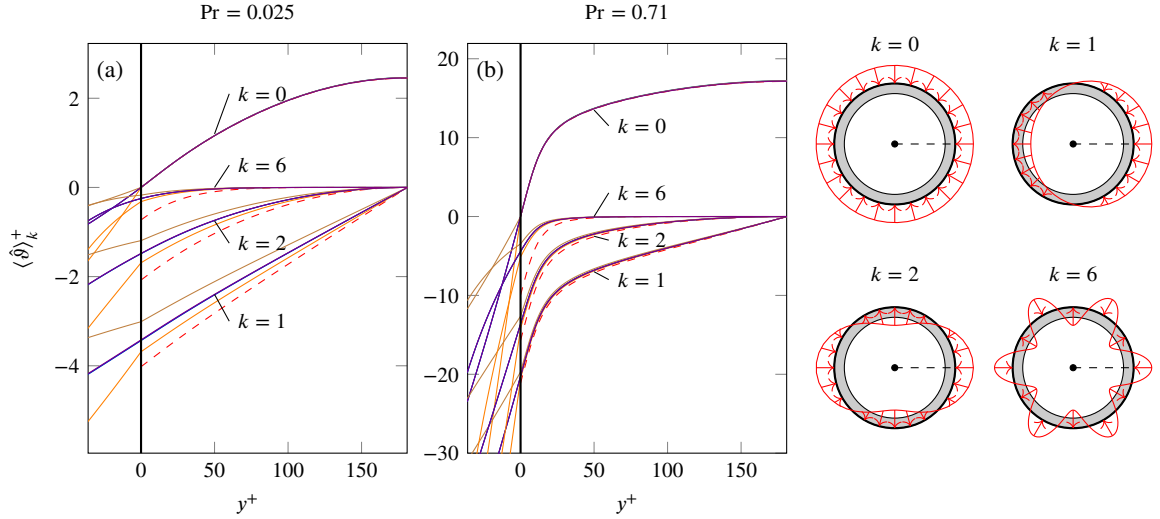

\paragraph*{Mean temperature.} \Cref{fig:mean_temp_modes} displays selected azimuthal temperature modes $\widehat{\avg{\vartheta}}_k$. The different line styles and colours correspond to solids of different properties as well as IF, with each cluster of lines annotated with its wavenumber $k$. For a single-wavenumber (i.e. monochromatic) boundary condition $f(\phi) = \Real{\exp{jk\phi}}$, the curve displayed in \cref{fig:mean_temp_modes} (that depends on $r$) multiplied with the boundary condition $f(\phi)$ is the resulting non-dimensional temperature distribution as a function of $r$ and $\phi$; hence $\avg{\hat{\vartheta}}_k$ is identified with the temperature profile (modulo a complex exponential) of a monochromatic boundary condition with appropriate heat flux scaling.

\begin{figure}
	\centering
	
	\input{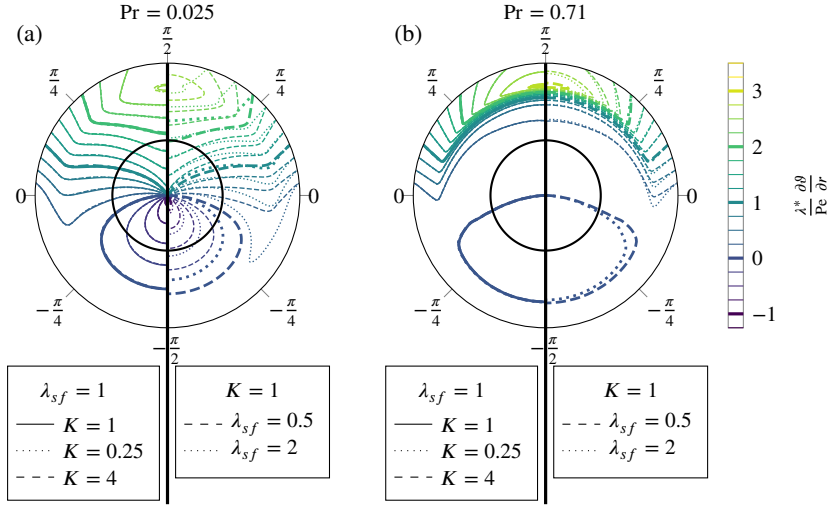}
	\caption{Comparison of $\pdv*{\vartheta}{r}$, normalised with $\lambda^* / \Pe$ for half-sinusoidal heat flux. Left side of each subplot: different $K$ for $\lambda_{sf} = 1$: 1~(solid), 0.25~(dotted), 4~(dashed). Right side of each subplot: different $\lambda_{sf}$ for $K=1$: 0.5~(dashed) and 2~(dotted).}
	\label{fig:halfsin_mean_dtdr}
\end{figure}

In \cref{sec:uniformhf}, it was shown that the presence of the solid has only minor influence of the fluid's mean temperature profile for uniform heat flux. This observation does not hold for azimuthally non-uniform temperature distributions (cf.~solid vs.~dashed lines for $k\neq 0$ in \cref{fig:mean_temp_modes}). Azimuthal conduction in the solid reduces the variation of the mean heat flux at the interface, thus reducing the magnitude of $\widehat{\avg{\vartheta}}_k$ within the fluid. This effect is particularly strong for large $k$ and small $\Pr$, which is readily explained. For increasing $k$, the characteristic length of azimuthal conduction is smaller, as the peaks of the boundary condition move closer together, and thus the impact of azimuthal conduction increases. For large $\Pr$, the solution is dominated by radial turbulent transport, so azimuthal conduction in the solid is less important in comparison to small-$\Pr$ fluids. As in the laminar case \citep{neuhauserSolutionExtendedGraetz2025}, a thick, well-conducting solid enhances this effect (cf.~brown (orange) curves in  \cref{fig:mean_temp_modes} for increased (reduced)  $\lambda_{sf}$). However, modifying $K$ only has a minor effect on the mean temperature profile also for $k > 0$: the purple, blue and green lines corresponding to different $K$ for fixed $\lambda_{sf}$ are indistinguishable in \cref{fig:mean_temp_modes}.
This behaviour readily carries over composite boundary conditions, exemplified by half-sinusoidal heat flux in  \cref{fig:halfsin_mean_dtdr}. In this figure, we display contours of radial heat flux ($\pdv*{\vartheta}{r}$). In the left semicircle of each panel, the contours of different $K$ at fixed $\lambda_{sf}$ are displayed. The contours coincide almost perfectly. In the right semicircles, $\lambda_{sf}$ is varied for constant $K$, and a pronounced influence of $\lambda_{sf}$ is observed. To summarize, while $K$ has minor influence on the in-plane mean temperature distribution for uniform and non-uniform heat flux, $\lambda_{sf}$ affects the effective boundary condition at the fluid-solid interface through azimuthal conduction in the solid, and affects the temperature distribution significantly in case of non-uniform boundary conditions (particularly for low-$\Pr$ fluids). 

One further observes in \cref{fig:mean_temp_modes} that for a given fluid-solid configuration and non-zero $k$, increasing $k$ reduces the magnitude of the modal temperature profile; and for high $k$ (cf. $k=6$ curves in \cref{fig:mean_temp_modes}) it quickly tends towards zero in the bulk of the flow. As in the laminar case (see \cite{neuhauserSolutionExtendedGraetz2025}), the bulk of the flow does not participate in the heat transfer for sufficiently large $k$; azimuthal (turbulent) diffusion already equilibrates temperature differences near (or even inside) the wall. This mechanism is observed for both $\Pr$ numbers considered, but generally more effective for higher $\Pr$. 

Finally, one observes in \cref{fig:mean_temp_modes} that for $k \geq 1$, $\avg{\hat{\vartheta}}_k|_{r=0} = 0$ by definition, and only the $k=1$ mode can have a non-zero gradient at the centreline for symmetry reasons. 

\paragraph*{Turbulent Prandtl number.} The direct solution of the statistically averaged temperature equation \cref{eq:rans_temperature} requires modelling of the terms $\avg{u_r'\vartheta'}$ and $\avg{u_\phi'\vartheta'}$. The gradient diffusion hypothesis \citep{popeTurbulentFlows2000} leads to the following closure (here in the dimensional form):
\begin{equation}
	\avg{u_r'T'} = - \alpha_{t,r}  \pdv{\avg{T}}{r},  \qquad \avg{u_\phi'T'} = - \alpha_{t,\phi} \frac{1}{r} \pdv{\avg{T}}{\phi}\label{eq:def_alphat}
\end{equation}
where $ \alpha_{t,r}$ and $ \alpha_{t,\phi}$ are the turbulent thermal diffusivities in the radial and azimuthal direction, respectively. They are often expressed in terms of the turbulent Prandtl number \citep{kaysTurbulentPrandtlNumber1994}:
\begin{equation}
	\Pr_{t,r} =  \frac{\nu_t}{\alpha_{t,r}}, \quad \Pr_{t,\phi} =  \frac{\nu_t}{\alpha_{t,\phi}}
\end{equation}
where $\nu_t$ is the turbulent eddy viscosity, here defined via the Boussinesq approximation $\avg{u_r'u_z'} = - \nu_t \pdv*{\avg{u_z}}{r}$. In common turbulent heat flux models such as the Kays correlation \citep{kaysTurbulentPrandtlNumber1994} or the $k{-}\epsilon{-}k_\theta{-}\epsilon_\theta$ model proposed by \citet{manservisiCFDFourParameter2014}, the turbulent Prandtl number is assumed scalar (i.e.\ $\Pr_{t,r} = \Pr_{t,\phi}$), although it is known that this is an inaccurate assumption especially near the wall \citep{blackExperimentsTurbulentHeat1967}. Previous studies \citep{antoranzperalesNumericalStudyTurbulent2017,straubAzimuthallyInhomogeneousThermal2019} have reported that for azimuthally non-uniform heating (of the half-sinusoidal type), the assumption of a scalar eddy diffusivity is clearly violated, even though isotropy is recovered near the pipe centre. 

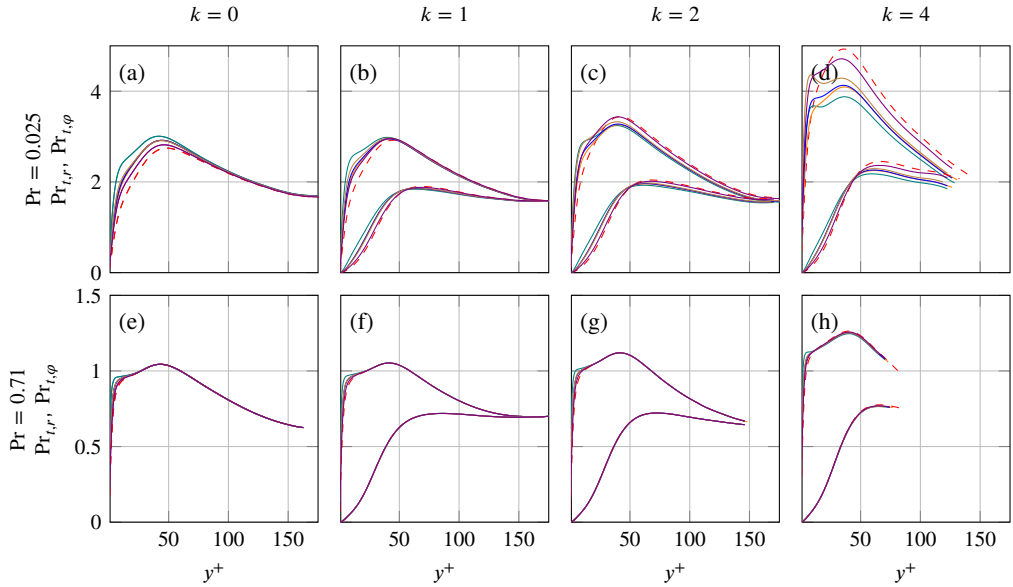
\begin{figure}
	\tikzsetnextfilename{prt_inhom_k}
	\centering
\begin{tikzpicture}
	\begin{groupplot}[group style={group size=4 by 2, horizontal sep=0.3cm, vertical sep=0.3cm,
    x descriptions at=edge bottom,
            y descriptions at=edge left}, xlabel=$y^+$, xmin=1, xmax=175, width=2.75cm, height=3cm, scale only axis=true, ymin=0, grid=both, ylabel style={align = center},  %
            ] 
    \nextgroupplot[ymax=5, ylabel={$\Pr = 0.025$ \\ $\Pr_{t,r}$, $\Pr_{t,\phi}$},subplotlabel=fig:mode_Prtr_0025_0, title={$k=0$}]	
    \plotHomTempProfileAllK{figdata/pipe/retau_180_les_r1.2_long/decomp_Pr_t_r_filtered_Pr0.025.csv}{0}{}
    \plotHomTempProfileAllK{figdata/pipe/retau_180_les_r1.2_long/decomp_Pr_t_phi_filtered_Pr0.025.csv}{0}{}
    \nextgroupplot[ymax=5, subplotlabel=fig:mode_Prtr_0025_1, title={$k=1$}]	
    \plotHomTempProfileAllK{figdata/pipe/retau_180_les_r1.2_long/decomp_Pr_t_r_filtered_Pr0.025.csv}{1}{}
    \plotHomTempProfileAllK{figdata/pipe/retau_180_les_r1.2_long/decomp_Pr_t_phi_filtered_Pr0.025.csv}{1}{}
    \nextgroupplot[ymax=5, subplotlabel=fig:mode_Prtr_0025_2, title={$k=2$}]	
    \plotHomTempProfileAllK{figdata/pipe/retau_180_les_r1.2_long/decomp_Pr_t_r_filtered_Pr0.025.csv}{2}{}
    \plotHomTempProfileAllK{figdata/pipe/retau_180_les_r1.2_long/decomp_Pr_t_phi_filtered_Pr0.025.csv}{2}{}
    \nextgroupplot[ymax=5, subplotlabel=fig:mode_Prtr_0025_3, title={$k=4$}]	
    \plotHomTempProfileAllK{figdata/pipe/retau_180_les_r1.2_long/decomp_Pr_t_r_filtered_Pr0.025.csv}{4}{}
    \plotHomTempProfileAllK{figdata/pipe/retau_180_les_r1.2_long/decomp_Pr_t_phi_filtered_Pr0.025.csv}{4}{}
	\nextgroupplot[ymax=1.5, subplotlabel=fig:mode_Prtr_071_0, ylabel={$\Pr = 0.71$ \\ $\Pr_{t,r}$, $\Pr_{t,\phi}$}]	
    \plotHomTempProfileAllK{figdata/pipe/retau_180_les_r1.2_long/decomp_Pr_t_r_filtered_Pr0.71.csv}{0}{}
    \plotHomTempProfileAllK{figdata/pipe/retau_180_les_r1.2_long/decomp_Pr_t_phi_filtered_Pr0.71.csv}{0}{}
    \nextgroupplot[ymax=1.5, subplotlabel=fig:mode_Prtr_071_1]	
    \plotHomTempProfileAllK{figdata/pipe/retau_180_les_r1.2_long/decomp_Pr_t_r_filtered_Pr0.71.csv}{1}{}
    \plotHomTempProfileAllK{figdata/pipe/retau_180_les_r1.2_long/decomp_Pr_t_phi_filtered_Pr0.71.csv}{1}{}
    \nextgroupplot[ymax=1.5, subplotlabel=fig:mode_Prtr_071_2]	
    \plotHomTempProfileAllK{figdata/pipe/retau_180_les_r1.2_long/decomp_Pr_t_r_filtered_Pr0.71.csv}{2}{}
    \plotHomTempProfileAllK{figdata/pipe/retau_180_les_r1.2_long/decomp_Pr_t_phi_filtered_Pr0.71.csv}{2}{}
    \nextgroupplot[ymax=1.5, subplotlabel=fig:mode_Prtr_071_5]	
    \plotHomTempProfileAllK{figdata/pipe/retau_180_les_r1.2_long/decomp_Pr_t_r_filtered_Pr0.71.csv}{4}{}
    \plotHomTempProfileAllK{figdata/pipe/retau_180_les_r1.2_long/decomp_Pr_t_phi_filtered_Pr0.71.csv}{4}{}
	\end{groupplot}
\end{tikzpicture}
\caption[Radial and azimuthal turbulent Prandtl number of selected azimuthal modes for $\Re_b = 5300$, varying $\Pr$ and different solid properties.]{Turbulent Prandtl number $\Pr_t$ in radial and azimuthal direction for selected modes at $\Re_b = 5300$. (a-d): $\Pr = 0.025$, (e-h): $\Pr = 0.71$. Columns from left to right: $k=\{0, 1, 2, 4\}$. In each subplot, the top cluster of curves correspond to $\Pr_{t,r}$ and the bottom cluster of curves correspond to $\Pr_{t,\phi}$. For $k=0$ (a, e), only $\Pr_{t,r}$ is relevant. Line styles correspond to different solid properties; see \cref{tab:pipe_cases_linestyles}.}
\label{fig:mean_prt_modes}
\end{figure}

In \cref{fig:mean_prt_modes}, the turbulent Prandtl numbers are presented for wavenumbers $k=\{0, 1, 2, 4\}$, the two rows of panels corresponding to different $\Pr$ and the columns to different $k$. The color and linestyle indicate solid properties or IF, as before. For $k \geq 1$, two clusters of lines are shown in each panels, where the top (bottom) cluster corresponds to $\Pr_{t,r}$ ($\Pr_{t,\phi}$). According to \cref{eq:def_alphat}, evaluation of $\alpha_{t,i}$ requires dividing by the temperature gradient, i.e.\ $\pdv*{\widehat{\avg{\vartheta}}_k}{r}$ and $\pdv*{\widehat{\avg{\vartheta}}_k}{\phi} = kj \widehat{\avg{\vartheta}}_k$. As shown in \cref{fig:mean_temp_modes}, both gradients vanish for $k > 1$ as $r \rightarrow 0$. Accordingly, the turbulent Prandtl number becomes ill-defined near the pipe centre and those regions are excluded from plotting. 

\Cref{fig:mean_prt_modes} confirms the observation of an anisotropic eddy diffusivity near the wall and a return to isotropy in the bulk for all $k>0$ and $\Pr$ considered, at least as far as it can be evaluated. In particular, $\Pr_{t,r} \sim y^+$ and $\Pr_{t,\phi} \sim (y^+)^2$ is observed for all wavenumbers at the wall. For $\Pr=0.71$, subfigures (e-h) show that $\Pr_{t,i}$ is virtually independent of the solid properties across all wavenumbers, except in a thin region near the wall. In contrast, for $\Pr=0.025$ (subfigures (a-d)) the solid properties influence $\Pr_{t,i}$ throughout the entire domain. For such low Prandtl numbers, instantaneous fluctuations of temperature near the wall (whose spectrum is strongly affected by the solid properties) are transported into the bulk primarily by molecular diffusion. This modifies the turbulent heat flux vector and, consequently, $\Pr_t$.

Both aforementioned studies observed that for a half-sinusoidal boundary condition, there is only a weak dependency of $\Pr_{t,i}$ on the azimuthal coordinate. If this statement held exactly, it would imply that $\Pr_{t,i}$ would be independent of the wavenumber $k$, which is not the case. However, $\Pr_{t,i}$ is indeed very similar for $k\leq 2$, which are the dominant modes of a half-sinusoidal boundary condition. $\Pr_{t,i}$ increases systematically with $k$, and this increase is particularly pronounced for the low-$\Pr$ case. This observation might be explained as follows: As the scalar field is driven at increasingly small azimuthal scales with growing $k$, azimuthal scalar gradients increase. This enhances molecular diffusion, which increases $\Pr_{t,i}$ in particular for low-$\Pr$ fluids where turbulent heat transfer is less dominant. At the same time, these azimuthal scales become too small to be effectively mixed by the dominant scales of turbulence, reducing the turbulent heat flux. Both effects contribute to the observed increase of $\Pr_{t,i}$ with $k$. One might foresee this to be an interesting test case for a turbulent heat flux model. 

\paragraph*{Temperature fluctuations.} 

\newcommand{\drawlegendcontourtwo}[1]{
	\draw (#1.south) node[below=0.5cm, left=0.1cm, draw=black,  anchor=north east] {\begin{tikzpicture}
			\draw (0.9, 0) node[above=0.15cm] {$K\vphantom{\lambda_{sf}}=1$};
			\draw (0, 0) -- ++ (0.5, 0) node[right] {$\lambda_{sf}=1$};
			\draw[dashed] (0, -0.35) -- ++ (0.5, 0) node[right] {$\lambda_{sf}=0.5$};
	\end{tikzpicture}};
	
	\draw (#1.south) node[below=0.5cm, right=0.1cm, draw=black, anchor=north west] {\begin{tikzpicture}
			\draw (0.75, 0) node[above=0.15cm] {$\lambda_{sf}=1$};
			\draw (0, 0) -- ++ (0.5, 0) node[right] {$K=1$};
			\draw[dashed] (0, -0.35) -- ++ (0.5, 0) node[right] {$K=4$};
	\end{tikzpicture}};}
\begin{figure}
	\pgfplotsset{
		keepright/.style={%
			x filter/.expression={%
				( x <= rad(90) ) || ( x >= rad(270) ) ? x : NaN%
			}%
		},
		keepleft/.style={%
			x filter/.expression={%
				( x >= rad(90) ) && ( x <= rad(270) ) ? x : NaN%
			}%
		}
	}
	\newcommand{\plotHomTempProfileAllInh}[1]{%
		
		\pgfplotstableread[col sep=comma]{#1}\datatable
		\pgfplotsinvokeforeach{IF,K0.25_L1.0,K1.0_L0.5,K1.0_L1.0,K1.0_L2.0,K4.0_L1.0}{%
			\addplot[
			mark=none,
			##1, %
			] table[x={y+}, y={##1_2}] from \datatable;
		}
	}
	\centering
	\tikzsetnextfilename{tempfluct_contour}
	\begin{tikzpicture}
		\begin{polaraxis}[at={(0, 0)}, yticklabel=\empty, xtick={0,45,...,359}, colormap/viridis, name=ax1, width=3.5cm, scale only axis, title={$\Pr=0.025$}, clip mode=individual,
			xticklabels={%
				$0$,
				$\frac{ \pi}{4}$,
				$\frac{ \pi}{2}$,
				$\frac{ \pi}{4}$,
				$0$,
				$-\frac{\pi}{4}$,
				$-\frac{\pi}{2}$,
				$-\frac{\pi}{4}$
			},ymax=0.6, ytick={0.5}, colorbar sampled line={scatter, 
			scatter/use mapped color={draw=mapped color,fill=mapped color},
			only marks, 
			mark=-, 
			mark size=0.1cm}, point meta min=0, point meta max=0.4, colorbar style={width=0.2cm, samples=9, axis y line*=box, ylabel=$\avg{\vartheta'\vartheta'}^+$},]
			\addplot[
			contour prepared={labels=false}, data cs = polarrad, 
			] table {figdata/pipe/retau_180_les_r1.2_long/tt_halfsin_Pr0.025_K1.0_L1.0_2.pgf};
			\addplot[
			contour prepared={labels=false}, data cs = polarrad, keepleft, dashed
			] table {figdata/pipe/retau_180_les_r1.2_long/tt_halfsin_Pr0.025_K1.0_L0.5_2.pgf};
			\addplot[
			contour prepared={labels=false}, data cs = polarrad, keepright, dashed,
			] table {figdata/pipe/retau_180_les_r1.2_long/tt_halfsin_Pr0.025_K4.0_L1.0_2.pgf};
			\draw[very thick] (90, 0.6) -- (270, 1.3);
			\addplot [
			thick, samples=180,
			domain=0:360,
			] gnuplot {0.5}				; 
		\end{polaraxis}
		\draw (ax1.outer north west) node[text width=1em,anchor=north west] {\subcaption{\label{plot:inhom:ttpr:contour:0.025}}};
		\drawlegendcontourtwo{ax1}
		
		\begin{polaraxis}[at={($(ax1.north east)+(3cm, 0cm)$)}, anchor=north west, yticklabel=\empty, xtick={0,45,...,359}, colormap/viridis, width=3.5cm, scale only axis,  name=ax2, title={$\Pr=0.71$}, clip mode=individual,
			xticklabels={%
				$0$,
				$\frac{ \pi}{4}$,
				$\frac{ \pi}{2}$,
				$\frac{ \pi}{4}$,
				$0$,
				$-\frac{\pi}{4}$,
				$-\frac{\pi}{2}$,
				$-\frac{\pi}{4}$
			},ymax=0.6, ytick={0.5}, colorbar sampled line={scatter, 
				scatter/use mapped color={draw=mapped color,fill=mapped color},
				only marks, 
				mark=-, 
				mark size=0.1cm}, point meta min=0, point meta max=60, colorbar style={width=0.2cm, samples=7, axis y line*=box, ylabel=$\avg{\vartheta'\vartheta'}^+$,},]
			\addplot[
			contour prepared={labels=false}, data cs = polarrad, 
			] table {figdata/pipe/retau_180_les_r1.2_long/tt_halfsin_Pr0.71_K1.0_L1.0_2.pgf};
			\addplot[
			contour prepared={labels=false}, data cs = polarrad, keepleft, dashed
			] table {figdata/pipe/retau_180_les_r1.2_long/tt_halfsin_Pr0.71_K1.0_L0.5_2.pgf};
			\addplot[
			contour prepared={labels=false}, data cs = polarrad, keepright, dashed,
			] table {figdata/pipe/retau_180_les_r1.2_long/tt_halfsin_Pr0.71_K4.0_L1.0_2.pgf};
			\draw[very thick] (90, 0.6) -- (270, 1.3);
			\addplot [
			thick, samples=180,
			domain=0:360,
			] gnuplot {0.5}				; 
		\end{polaraxis}
		\draw (ax2.outer north west) node[text width=1em,anchor=north west] {\subcaption{\label{plot:inhom:ttpr:contour:0.71}}};
		\drawlegendcontourtwo{ax2}

		\begin{groupplot}[group style={group size=2 by 3, }, width=5cm,
			height=4cm,
			xmin=-36,
			xmax=180,
			scale only axis,
			grid=both,
			]
			\nextgroupplot[clip=true, ylabel={$\overline{\avg{\vartheta'\vartheta'}^+}$}, ymin=0, subplotlabel=plot:inhom:ttpr:mean:0.025, at={($(ax1.south)+(0, -2.5cm)$)}, anchor=north]
			\draw[thick] (0, \pgfkeysvalueof{/pgfplots/ymin}) -- (0, \pgfkeysvalueof{/pgfplots/ymax}) node[above right] {fluid} node[above left] {solid};
			\plotHomTempProfileAllInh{figdata/pipe/retau_180_les_r1.2_long/inhom_trms_profile_Pr_0.025.csv}

			\nextgroupplot[clip=true, ymin=0, xmax=80, subplotlabel=plot:inhom:ttpr:mean:0.71,  at={($(ax2.south)+(0, -2.5cm)$)}, anchor=north]
			\draw[thick] (0, \pgfkeysvalueof{/pgfplots/ymin}) -- (0, \pgfkeysvalueof{/pgfplots/ymax}) node[above right] {fluid} node[above left] {solid};
			\plotHomTempProfileAllInh{figdata/pipe/retau_180_les_r1.2_long/inhom_trms_profile_Pr_0.71.csv}
			
		\end{groupplot}
	\end{tikzpicture}
	\caption[Axial profile and azimuthal mean of $\avg{\vartheta'\vartheta'}$ for $\Re_b = 5300$, varying $\Pr$ and different solid properties (half-sinusoidal heat flux).]{Variance of $\vartheta$ for half-sinusoidal heating, $\Re_b = 5300$. The left column ((a), (c)) displays $\Pr=0.025$, the right column ((b), (d)) $\Pr=0.71$. Top row (a-b): Contour lines of $\avg{\vartheta'\vartheta'}^+$, comparing the effect of varying solid properties (line style). Solid lines: $K=\lambda_{sf} = 1$. Dashed lines in each panel: Left semicircle: $K=1$, $\lambda_{sf} = 0.5$, right semicircle: $K=4$, $\lambda_{sf} = 1$. Bottom row (c-d): Azimuthally averaged temperature variance. Line styles correspond to different solid properties; see \cref{tab:pipe_cases_linestyles}.}
	\label{fig:tempfluct_inhom}
\end{figure}
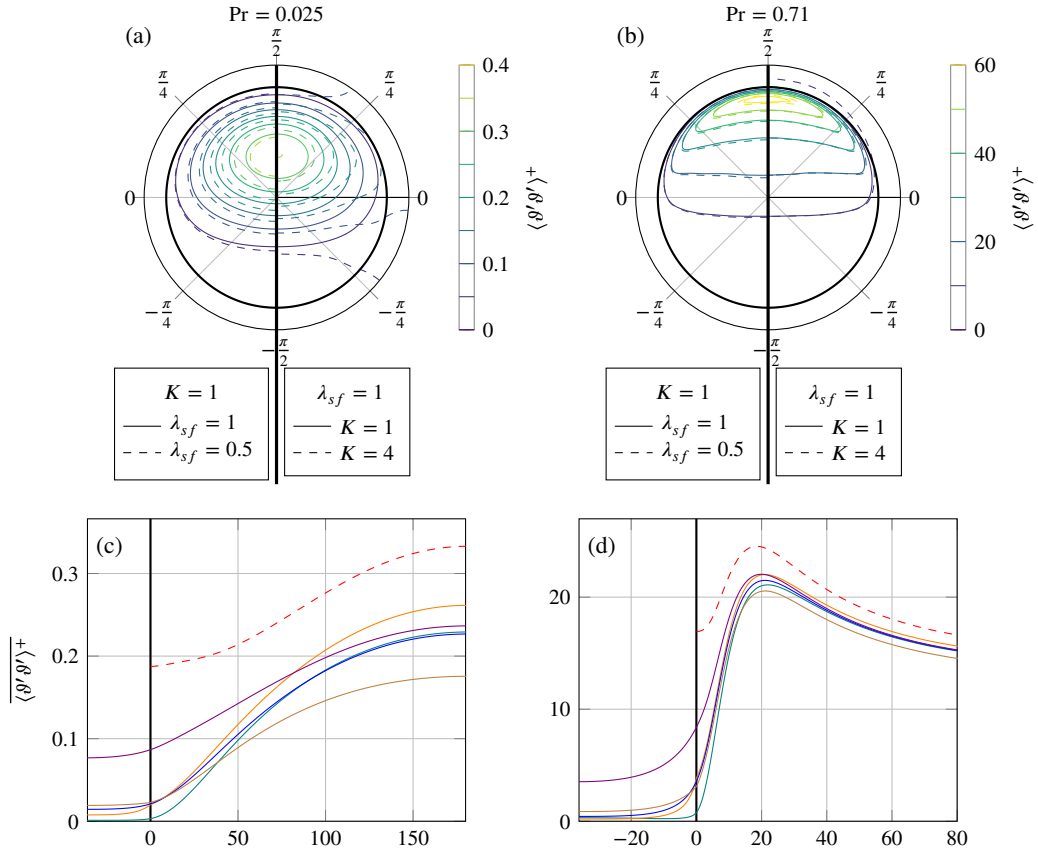

Unlike for first-order statistics, nonlinear effects prohibit a convenient modal analysis of the temperature variance $\avg{\vartheta'\vartheta'}$ for arbitrary boundary conditions. Instead, \cref{plot:inhom:ttpr:contour:0.025,plot:inhom:ttpr:contour:0.71} display cross-sectional profiles of $\avg{\vartheta'\vartheta'}$ for the half-sinusoidal heating case; linestyles indicate different solid properties as indicated in the legend of the panels.  For both $\Pr$, varying $\lambda_{sf}$ for constant $K$ (left semicircle of each subplot) has minor influence on $\avg{\vartheta'\vartheta'}$. Increasing $K$ for constant $\lambda_{sf}$ (right semicircles) allows fluctuations to travel further into the solid. The global peak of temperature variance remains at $\phi = \pi / 2$ for all considered solid properties, and the general shape of the fluctuation profile -- solid and near-wall region excluded -- is consistent between cases.

For a quantitive comparison of the fluctuation, \cref{plot:inhom:ttpr:mean:0.025,plot:inhom:ttpr:mean:0.71} present the azimuthally averaged temperature variance as a function of wall distance with the line style / colour as in \cref{tab:pipe_cases_linestyles}, and for both $\Pr$ numbers (columns). \citet{straubAzimuthallyInhomogeneousThermal2019} presented data for the same setup without conjugate heat transfer. As in the case of uniform heating, increasing the effusivity ratio $K$ amplifies temperature fluctuations within the solid; this effect is particularly pronounced for low $\Pr$. The temperature variance near the centerline is only weakly affected by $K$ (compare purple, blue and green lines in \cref{plot:inhom:ttpr:mean:0.025,plot:inhom:ttpr:mean:0.71}). 

Unlike for uniform heat flux, the conductivity ratio $\lambda_{sf}$ has a strong impact on temperature fluctuations. A thick, well conducting wall reduces mean temperature gradients at the interface in both radial and azimuthal directions, directly decreasing the production of temperature variance within the fluid, which is (in general) given by 
\begin{equation}P_{\vartheta'\vartheta'} = \avg{\vartheta' u_r'} \pdv{\avg{\vartheta}}{r} + \frac{\avg{\vartheta' u_\phi'}}{r} \pdv{\avg{\vartheta}}{\phi} + \avg{\vartheta' u_z'} \pdv{\avg{\vartheta}}{z}.\end{equation}
The reduction of mean gradients due to solid diffusivity is, as discussed, significantly more effective for small $\Pr$. As $\lambda_{sf}$ increases (brown vs. blue line in \cref{plot:inhom:ttpr:mean:0.025,plot:inhom:ttpr:mean:0.71}), temperature fluctuations in the fluid are reduced, while they are amplified in the solid. These observations have a direct engineering implication for applications employing low-$\Pr$ fluids under non-uniform heating conditions. By adjusting the wall thickness and/or solid properties, one can substantially reduce the variance of temperature fluctuations in the fluid, thereby making more efficient use of the heat transfer fluid while keeping it within safe operating limits, at the cost of slightly increasing temperature fluctuations in the solid and thus, potentially, thermal stresses. Such modifications leaves the global heat-transfer coefficient essentially unchanged (see discussion of \cref{fig:meantemp_pr071}).

\citet{straubAzimuthallyInhomogeneousThermal2019} observed that for the same $\Re_b$ and $\Pr$, half-sinusoidal heating instead of uniform heat flux increases the maximum value of the temperature variance by approximately a factor of 10. This can also be observed in \cref{plot:inhom:ttpr:contour:0.025,plot:inhom:ttpr:contour:0.71} relative to \cref{fig:trms_pr0025,fig:trms_pr071}. For $\Pr=0.71$, the maxima of wall heat flux and wall temperature variance coincide, whereas for $\Pr=0.025$ they are offset by approx.~$\pi / 4$ in either direction. Motivated by this observation, a simple model for the temperature variance production can be constructed based on truncating the modal representation. First, inserting the definition of the turbulent eddy diffusivity \cref{eq:def_alphat} into the definition of temperature variance production and observing from \cref{fig:mean_prt_modes} that the turbulent eddy diffusivity is similar for the first (and for cases of half-sided heating typically dominant) wavenumbers $k=0, 1, 2$, one has
\begin{align*}
	P_{\vartheta'\vartheta'} &= \avg{\vartheta' u_r'} \pdv{\avg{\vartheta}}{r} + \frac{\avg{\vartheta' u_\phi'}}{r} \pdv{\avg{\vartheta}}{\phi} + \avg{\vartheta' u_z'} \pdv{\avg{\vartheta}}{z} = - \alpha_{t,r}  \left(\pdv{\avg{\vartheta}}{r}\right)^2 - \frac{\alpha_{t,\phi}}{r^2} \left(\pdv{\avg{\vartheta}}{\phi}\right)^2 \\
	&\approx - \alpha_{t,r}  \left(\pdv{\widehat{\avg{\vartheta}}_0}{r} + \frac{\pi}{2} \sin{\phi} \pdv{\widehat{\avg{\vartheta}}_1}{r} - \frac{2}{3} \cos{2\phi} \pdv{\widehat{\avg{\vartheta}}_2}{r} + ... \right)^2 - \frac{\alpha_{t,\phi}}{r^2} \left(\frac{\pi}{2} \cos{\phi} \widehat{\avg{\vartheta}}_1 + \frac{4}{3} \sin{2\phi} \widehat{\avg{\vartheta}}_2 + ... \right)^2
\end{align*}
where the numerical coefficients ($1$, $\pi/2$, $2/3$) arise from the Fourier representation of the half-sinusoidal boundary condition, truncated after $k=2$. Finally, \cref{fig:mean_temp_modes} indicated that, in the outer half of the pipe (where the peak of temperature variance production is located), the magnitudes of the radial gradients of the first three modes are comparable, i.e.\ $\mathcal{O}(\pdv*{\widehat{\avg{\vartheta}}_0}{r}) = \mathcal{O}(\pdv*{\widehat{\avg{\vartheta}}_1}{r}) = \mathcal{O}(\pdv*{\widehat{\avg{\vartheta}}_2}{r})$, and that $\widehat{\avg{\vartheta}}_1 \approx  2 \widehat{\avg{\vartheta}}_2$. Inserting, one finds
\begin{align}
	P_{\vartheta'\vartheta'} &\approx
	- \alpha_{t,r} \left(\pdv{\widehat{\avg{\vartheta}}_0}{r}\right)^2 \left(1 + \frac{\pi}{2} \sin{\phi} - \frac{2}{3} \cos{2\phi} \right)^2 - \frac{\alpha_{t,\phi}}{r^2} \left(\widehat{\avg{\vartheta}}_1\right)^2 \left(\frac{\pi}{2} \cos{\phi}  + \frac{2}{3} \sin{2\phi} \right)^2 \notag\\
	&\approx - \alpha_{t,r} \left(\pdv{\widehat{\avg{\vartheta}}_0}{r}\right)^2 f(\phi)^2  - \frac{\alpha_{t,\phi}}{r^2} \left(\widehat{\avg{\vartheta}}_1\right)^2 f'(\phi)^2 \label{eq:model_production_of_tt}
\end{align}
For half-sinusoidal heating, $f(\phi)^2$ attains its maximum value $\pi^2 \approx 9.86$ at $\phi = \pi/2$, while $f'(\phi)^2$ has the same maximum at the inflection points of $f(\phi)$, i.e.\ at $\phi = 0$ and $\phi =  \pi$. The Prandtl number alters the relative importance of the azimuthal and radial production terms: increasing $\Pr$ enhances $\Nu_l$ and thus increases radial production (first term). For $y^+ \lesssim 20$, the near-wall behaviour implies $\alpha_{t,\phi} \gg \alpha_{t,r}$ (see discussion of $\Pr_t$, \cref{fig:mean_prt_modes}), so the azimuthal production (second term) is locally amplified near the wall. To summarise, a production peak is expected at $\phi = \pi/2$, while for low $\Pr$, secondary near-wall peaks are expected offset from $\pi/2$ by $\Delta \phi \leq \pi/2$ -- the actual position is determined by the balance of azimuthal and radial contributions to the production, and indeed it varies slightly between cases of different solid properties. Interestingly, the low-order model captures nicely the peak locations and $\mathcal{O}(10)$ increase of temperature variance compared to the uniform heat flux case, which was also observed by \citet{straubAzimuthallyInhomogeneousThermal2019} -- even though the precise level of temperature variance is not only governed by its production, but also dissipation and (turbulent and molecular) diffusion.

\section{Conclusions}

This paper investigates CHT in thermally and hydraulically fully developed turbulent pipe flow subject to azimuthally uniform and non-uniform thermal boundary conditions. A detailed analysis of the spectral structure of the system provides a framework for treating arbitrary azimuthal heat-flux distributions. For first-order temperature moments, the analysis reduces the parameter space from all sufficiently regular boundary conditions $f(\phi)\in L^2$ to the space of positive azimuthal wavenumbers. It also determines the azimuthal phase of quantities such as $\avg{\vartheta}$, $\avg{\vartheta'u_r'}$, and $\avg{\vartheta'u_\phi'}$. For higher-order moments, the same framework constrains the azimuthal spectrum and is later used to estimate the production of temperature fluctuations.

An important consequence of the spectral analysis concerns the global heat-transfer performance. Azimuthal means of first-order temperature moments are shown to depend only on the mean heat flux and not on its variation along the circumference. Consequently, the global Nusselt number $\Nu$ is independent of the azimuthal heat-flux distribution. This resolves the open question raised by \citet{maroccoNumericalAnalysisSolar2016} of whether the equivalence previously observed for conventional fluids also applies at low Prandtl numbers: within the passive-scalar approximation, the result is independent of $\Pr$ and follows from linearity and rotation equivariance of the temperature equation and its boundary conditions. Hence, $\Nu$ independence on spatial variations of the boundary condition holds for a much wider class of passive-scalar turbulent heat transfer problems, such as in plane channel flow, boundary layer flow, spatially or temporally developing flow and axially rotating pipe flow. 

The theoretical analysis is complemented by a well-resolved numerical dataset obtained with the spectral-element solver NekRS at $\Re_\tau \approx 180$. A relatively long domain of $L_z=100R_i$ is employed to avoid underestimation of temperature fluctuations at low Prandtl numbers. Various thermal boundary conditions and solid properties are considered at $\Pr\in\{0.025,0.71\}$, representative of liquid metals and air, respectively. Azimuthally uniform heat flux serves as a baseline configuration and aligns well with literature results. We first investigate the non-dimensionalisation with the effusivity ratio $K$ and the thermal-conductivity ratio $\lambda_{sf}$. While $K$ has little influence on the mean temperature field, it governs the level of temperature fluctuations. The mixed boundary condition (MBC) and isoflux (IF) cases emerge as limiting cases for $K\rightarrow0$ and $K\rightarrow\infty$, respectively, consistent with the behaviour previously reported by \citet{tiseljDNSTurbulentHeat2001} for turbulent channel flow. Overall, conjugate heat transfer has only a minor influence on the statistically averaged temperature field in the fluid region and on $\Nu$. 

For azimuthally non-uniform heat flux, a test boundary condition is used to determine simultaneously the response of first-order temperature moments for wavenumbers $k=0,\ldots,10$. In contrast to the uniform mode $k=0$, the modal mean-temperature distribution for $k>0$ depends strongly on the solid properties because heat conduction in the solid modifies the effective thermal boundary condition seen by the fluid. With increasing $k$, the response becomes progressively confined to the solid and the near-wall fluid region, while the pipe core increasingly ceases to participate in the heat transfer. 

Based on the generated DNS data, we evaluate the radial and azimuthal turbulent Prandtl numbers. For every $k>0$, $\Pr_t$ is anisotropic, with $\Pr_{t,r}>\Pr_{t,\phi}$ near the wall and a return towards isotropy in the pipe centre. The turbulent Prandtl number generally increases with $k$, but only moderately for $k\leq2$, which are the dominant modes of a half-sinusoidal heat flux distribution. This explains why \citet{straubAzimuthallyInhomogeneousThermal2019,antoranzperalesNumericalStudyTurbulent2017} observed only a weak azimuthal variation of the turbulent Prandtl number for this boundary condition. The influence of the solid properties on $\Pr_{t,i}$ also depends strongly on the fluid Prandtl number: for $\Pr=0.025$, it extends across the pipe, whereas for $\Pr=0.71$ it is largely confined to the near-wall region.

Finally, higher-order moments are illustrated by example through the temperature variance under half-sinusoidal heating. \citet{straubAzimuthallyInhomogeneousThermal2019} previously reported an approximately tenfold increase in temperature variance relative to uniform heating, when non-dimensionalised with the mean heat flux. A simple production model based on the spectral analysis quantitatively reproduces this increase and associates local maxima of temperature variance with extrema of the imposed boundary condition (high radial temperature gradients), and with its inflection points (high azimuthal temperature gradients). Under conjugate heat-transfer conditions, increasing the solid conductivity reduces temperature fluctuations in the fluid, which is favourable for limiting fluid-temperature excursions, but increases them in the solid, with corresponding implications for thermal stresses.

The present investigation is restricted to thermally and hydraulically fully developed conditions and to $\Re_\tau\approx180$. Confirming the observed behaviour, in particular that of the temperature variance and turbulent Prandtl number, at higher Reynolds numbers is an important next step. A further extension concerns axially developing conjugate heat transfer. We have previously considered axial development under azimuthally non-uniform boundary conditions in \citet{iTi2025} and we expect that coupling such development to conjugate heat transfer will substantially increase the thermal development length, particularly at high Péclet numbers.

\appendix
\crefalias{section}{appendix}

\section{Spectral structure of higher-order moments of temperature}
\label{app:higher-order}

Statistical moments of temperature are given by \citet{popeTurbulentFlows2000} for $m, n \geq 0$ as
\begin{equation}
	\begin{split}
		&\avg{u_i{}^m \vartheta{}^n}(r, \phi, z, t) =   \iiint_{-\infty}^\infty \int_{-\infty}^\infty u_i{}^m \vartheta{}^n p(u_r, u_\phi, u_z, \vartheta, r, \phi, z, t) \dd \vartheta \dd u_r \dd u_\phi \dd u_z. 
	\end{split}\label{stat_moment}
\end{equation}
Azimuthal integration of $\avg{u_i{}^m \vartheta{}^n}$, weighted with $\exp{j l \phi}$, yields the contribution of the mode corresponding to wavenumber $l$; in particular for $l=0$, $\frac 1{2\pi} \int_0^{2\pi} e^{j l \phi} \avg{u_i{}^m \vartheta^n} \dd \phi = \overline{\avg{u_i{}^m \vartheta{}^n}} (r, z, t)$. 

From linearity and rotation equivariance, one can show \cite{iTi2025}
\begin{equation}
	p(u_r, u_\phi, u_z, - \vartheta, r, \phi + \pi / k, z, t) = p(u_r, u_\phi, u_z, \vartheta, r, \phi, z, t). \label{pdf-transformation}
\end{equation}
This antisymmetry in the joint velocity-temperature PDF can be used to infer useful restrictions on higher-order statistical moments; the argument is similar to showing that the $r{-}z$ and $r{-}\phi$ components of the Reynolds stress tensor must be zero \citep{popeTurbulentFlows2000}. The moment $\avg{u_i{}^m \vartheta{}^n}$ then has a weighted mean of $\frac 1{2\pi} \int_0^{2\pi} \exp{j l \phi} \avg{u_i{}^m \vartheta{}^n}(r, \phi, z, t) \dd \phi = \iiint_{-\infty}^\infty  B \dd u_r \dd u_\phi \dd u_z$, where $B$ is given by
\begin{align*}
	B &:= \frac 1 {2\pi} \int_0^{2\pi} \exp{j l \phi} \int_{-\infty}^\infty   u_i{}^m \vartheta{}^n p(u_j, \vartheta, r, \phi, z, t) \dd \vartheta \dd \phi  \\
	&= \sum_{a = 0}^{k-1} \frac 1 {2\pi} \int_0^{2\pi / k} \exp{j l (\phi +2 \pi a/ k) } 
	\int_{-\infty}^\infty   u_i{}^m \vartheta{}^n p(u_j, \vartheta, r, \phi, z, t) \dd \vartheta \dd \phi \\
	&= \left(\sum_{a = 0}^{k-1} \left(\exp{2 \pi j l /k}\right)^a\right) \frac 1 {2\pi} \int_0^{2\pi / k} e^{j \phi l} 
	\int_{-\infty}^\infty u_i{}^m \vartheta{}^n p(u_j, \vartheta, r, \phi, z, t) \dd \vartheta \dd \phi 
\end{align*}
The geometric sum $S = \sum_{a = 0}^{k-1} w^a$ with $w =\exp{2 \pi j l /k} $ is readily evaluated; if $l$ is a multiple of $k$ (i.e.\ $l = \alpha k$ with integer $\alpha $), then $w=1$ and $S=k$. Otherwise, $S =  \frac{1 - e^{2 \pi j l}}{1 - e^{2 \pi j l /k}} = 0$ as the numerator is zero and the denominator is nonzero. For $l = \alpha k$ one can split the  remaining integral at $\pi / k$, substituting $\phi \rightarrow \phi + \pi/k$ and $\vartheta \rightarrow -\vartheta$ and finally using \cref{pdf-transformation}:
\begin{align*}
	B &=  \frac {k} {2\pi} \int_0^{2\pi / k} \exp{j \phi \alpha k} 
	\int_{-\infty}^\infty   u_i{}^m \vartheta{}^n p(u_j, \vartheta, r, \phi, z, t) \dd \vartheta \dd \phi \\ 
	&= \frac {k} {2\pi} \Biggl(\int_0^{\pi / k} \exp{j \phi \alpha k} 
	\int_{-\infty}^\infty   u_i{}^m \vartheta^n p(u_j, \vartheta, r, \phi, z, t) \dd \vartheta \dd \phi \\
	& \qquad \qquad + \int_{\pi/k}^{2\pi / k} \exp{j \phi \alpha k} 
	\int_{-\infty}^\infty   u_i{}^m \vartheta{}^n p(u_j, \vartheta, r, \phi, z, t) \dd \vartheta \dd \phi \Biggr) \\
	&= \frac {k} {2\pi} \Biggl(\int_0^{\pi / k} \exp{j \phi \alpha k} 
	\int_{-\infty}^\infty   u_i{}^m \vartheta{}^n p(u_j, \vartheta, r, \phi, z, t) \dd \vartheta \dd \phi \\
	& \qquad \qquad + \int_0^{\pi / k} \exp{j (\phi + \pi/k) \alpha k} 
	\int_{-\infty}^\infty   u_i{}^m \vartheta{}^n p(u_j, \vartheta, r, \phi + \pi / k, z, t) \dd \vartheta \dd \phi \Biggr) \\
	&= \frac {k} {2\pi} \left(1 + (-1)^n \exp{j\pi \alpha }\right) \int_0^{\pi / k} e^{j \phi \alpha k} 
	\int_{-\infty}^\infty   u_i{}^m \vartheta{}^n p(u_j, \vartheta, r, \phi, z, t) \dd \vartheta \dd \phi
\end{align*}
The combinations ($\alpha$ odd, $n$ even) and ($\alpha$ even, $n$ odd) lead to $B = 0$.\footnote{Clearly, $B=0$ for $n=1$ and $\alpha=0$, i.e.\ boundary conditions with $k\neq 0$ lead to first-order temperature statistics with vanishing azimuthal mean.}. Hence, for $f(\phi) = \sin(k \phi)$, the only remaining modes in $\avg{\vartheta^2}$ are $l \in \{0, 2k, 4k, 6k, \dots \}$ and for $\avg{\vartheta^3}$ those of $l \in \{k, 3k, 5k,  \dots \}$.

\section*{Acknowledgements}
We greatly acknowledge the support by the German Research Foundation (DFG) under research project 423710075. The simulations were performed on the national supercomputer HPE Cray EX4000 Hunter at the High Performance Computing Center Stuttgart (HLRS) under the grant number ctbctpf.

\section*{Data availability}

Mean statistics of the presented numerical simulations are available in \cite{neuhauserDatasetConjugateHeat2026}.

\printcredits

\bibliographystyle{cas-model2-names}

\bibliography{literature}

\end{document}